\documentclass[sigconf,balance=false]{acmart}
\usepackage{popets}
\usepackage{balance}

\setcopyright{popets}
\copyrightyear{YYYY}

\acmYear{YYYY}
\acmVolume{YYYY}
\acmNumber{X}
\acmDOI{XXXXXXX.XXXXXXX}
\acmISBN{}
\acmConference{Proceedings on Privacy Enhancing Technologies}
\usepackage{xspace}
\usepackage{hyperref}
\usepackage{verbatim}
\newcommand{\sysname}{QUICstep\xspace}

\newcommand{\channelname}{handshake channel\xspace}
\newcommand{\qsproxy}{handshake channel provider\xspace}

\usepackage[T1]{fontenc}

\usepackage{color}

\usepackage{graphicx}

\usepackage[labelformat=simple]{subcaption}
\usepackage{xspace}
\usepackage{multirow}
\usepackage{makecell} 
\usepackage{enumitem} 

\usepackage{ulem}
\usepackage{listings}
\usepackage[export]{adjustbox}
\usepackage{multirow}
\usepackage{booktabs}

\usepackage{algpseudocode}
\usepackage{algorithm}

\usepackage{bigdelim}
\usepackage{changepage}
\usepackage{float}
\usepackage{multicol}

\newcommand{\figref}[1]{Figure~\ref{#1}}
\newcommand{\tableref}[1]{Table~\ref{#1}}
\newcommand{\secref}[1]{\S\ref{#1}}

\ifdefined\isFinalized
\newcommand{\NewCommentType}[3]{}
\else
\newcommand{\NewCommentType}[3]{\expandafter\newcommand\csname #1\endcsname[1]{{\color{#2}{#3: ##1}} }}
\fi

\usepackage[override]{cmtt} 

\newcommand{\sjl}[1]{{\textcolor{black}{#1}}}
\newcommand{\cam}[1]{{\textcolor{black}{#1}}}

\begin{document}

\title[QUICstep: Evaluating connection migration based QUIC censorship circumvention]{QUICstep: Evaluating connection migration based QUIC censorship circumvention}


\author{Seungju Lee} 
\email{seungjulee@princeton.edu}
\affiliation{%
  \institution{Princeton University}
  \city{}
  \state{}
  \country{}
}

\author{Mona Wang} 
\email{monaw@princeton.edu}
\affiliation{%
  \institution{Princeton University}
  \city{}
  \state{}
  \country{}
}

\author{Watson Jia} 
\email{watsonj@alumni.princeton.edu}
\affiliation{%
  \institution{Princeton University}
  \city{}
  \state{}
  \country{}
}

\author{Qiang Wu} 
\email{gfw.report@protonmail.com}
\affiliation{%
  \institution{GFW Report}
  \city{}
  \state{}
  \country{}
}

\author{Henry Birge-Lee} 
\email{birgelee@princeton.edu}
\affiliation{%
  \institution{Princeton University}
  \city{}
  \state{}
  \country{}
}

\author{Liang Wang} 
\email{lw19@princeton.edu}
\affiliation{%
  \institution{Princeton University}
  \city{}
  \state{}
  \country{}
}

\author{Prateek Mittal} 
\email{pmittal@princeton.edu}
\affiliation{%
  \institution{Princeton University}
  \city{}
  \state{}
  \country{}
}



\begin{abstract}
Internet censors often rely on information in the first few packets of a connection to censor unwanted traffic. 
With the rise of the QUIC transport protocol, prior work has suggested the method of using QUIC \emph{connection migration} to conceal the first few handshake packets using a different network path (e.g., an encrypted proxy channel).
However, the use of connection migration for censorship circumvention has not been explored or validated in terms of feasibility or performance.
We bridge this gap by providing a rigorous \textcolor{black}{quantitative} evaluation of this approach that we name \sysname.
We develop a lightweight, application-agnostic prototype of \sysname and demonstrate that \sysname is able to circumvent a real-world QUIC SNI censor. 
We find that not only does 
\sysname outperform a fully encrypted channel in diverse settings, but also that \textcolor{black}{it can significantly reduce traffic load for encrypted channel providers}.
We also propose using \sysname as a tool for measuring QUIC connection migration support in the wild and show that support for connection migration is on the rise.
\textcolor{black}{While as of now QUIC and connection migration support is limited, we envision that \sysname can be a useful tool for the future where QUIC is the de facto norm for the Internet.}
\end{abstract}


\maketitle

\section{Introduction} \label{sec:intro}


As the Internet has become an indispensible tool in our lives, governments concurrently seek to expand censorship programs in a race to control our access to information. Access Now's 2024 data shows that network disruptions are not only occurring more frequently, but also across more countries~\cite{accessnow}.
The proliferation of commodity network devices that can perform deep packet inspection (DPI) has made scalable network censorship available to a wider range of governments and ISPs~\cite{sundararaman2020measuring}.
Well-established censorship regimes, such as China's Great Firewall (GFW) and Russia's TSPU, continue to deepen their methods for blocking network connections and identifying circumvention technologies~\cite{Hoang2021a, Hoang2024a, xue2022tspu}.

From previous studies of censorship systems in the wild, the vast majority of connections, including fully encrypted connections, are filtered based on information in the handshake packets.
For example, SNI-based detection is so critical to censorship~\cite{Chai2019a} that censors began prematurely blocking connections using the encrypted SNI extension as early as 2020~\cite{2020Bock}.
Similarly, the GFW only prioritizes the first few packets of a connection when deciding whether to exempt it from blocking~\cite{Wu2023a}.
Censorship measurement platforms such as OONI and Censored Planet leverage this to measure censorship at a global scale by primarily sending handshake probes~\cite{ooni,censored_planet}.
Ultimately, the handshake is critical in providing censors with sufficient connection metadata to make a judgment on whether to block the rest of the connection.

Hiding the handshake from the censor or bootstrapping the connection via another channel is key to circumventing handshake-dependent censorship in practice.
In this line, Wang et al. have suggested that connection migration can be used to circumvent stateless censorship of QUIC by splitting handshake and non-handshake packets across different network paths~\cite{wang2022leveraging}.
Connection migration is a notable feature of QUIC that allows connections to be maintained while the IP address or port of endpoints change.
\textcolor{black}{By transmitting only handshake packets through a censorship-resilient channel (e.g. VPN), users can minimize the performance overhead and load incurred by the censorship-resilient channel
while enjoying the benefits of circumventing censorship. }

However, prior work only briefly suggests this censorship circumvention scheme and leaves unanswered critical questions about its design, implementation, practicality, and \textcolor{black}{quantitative} performance \textcolor{black}{benefits}.

\paragraph{Contributions.} We name this approach \sysname and seek to understand its feasibility and performance: \textbf{First, are web servers compatible with \sysname? Second, can \sysname effectively circumvent deployed censors? Third, what is the \textcolor{black}{quantitative} performance impact of using  
\sysname?} 
To address these questions, we first define a clear threat model~(\secref{sec:threatmodel}) 
that focuses on the behavior of deployed censors that use simple and lightweight mechanisms (e.g., stateless behavior) over comprehensive but complex mechanisms. We then develop a lightweight, open-source prototype of \sysname~(\secref{sec:implementation}) that enables efficient path migration in a real-world web browsing setting.
Our implementation is available at \cam{\url{https://github.com/inspire-group/QUICstep}}.
Using this implementation, we perform comprehensive evaluations to demonstrate the effectiveness and performance of \sysname.

We successfully use \sysname to circumvent active QUIC-SNI censorship in the wild, showcasing its effectiveness against real-world deployed censors who leverage stateless detection techniques (\secref{sec:real-world-censor}). While more sophisticated censors could employ stateful traffic analysis to detect \sysname, this may increase censor's cost; we discuss this possibility and tradeoff in a more thorough security analysis of \sysname (\secref{sec-analysis}). 

We demonstrate the performance benefits of \sysname across a wide range of experiments. We use the term \emph{\channelname} to describe the secure tunnel through which a \sysname client performs the handshake~(\secref{design:client}).
Compared to \textcolor{black}{completely relying on the \channelname for all traffic}, 
\sysname can reduce page load time by up to 84\%, and load on \channelname providers by a median of \textbf{93\%}. The performance gain of \sysname becomes more pronounced when the \channelname is operating under practical limitations, such as bandwidth limits. \textcolor{black}{\sysname could be used to reduce load on \channelname providers, such as VPNs or other censorship circumvention tools.} 


For \sysname to be practically useful, the website must support both QUIC and connection migration, which currently presents a bottleneck for widespread use of such techniques. Leveraging \sysname, we conduct a large-scale measurement of connection migration support in the wild over a three-month period, identifying a small but nontrivial number of QUIC websites that are compatible with \sysname. 
\textcolor{black}{While support for both QUIC and connection migration is currently limited, } the support is also on the rise: the number of websites that partially support connection migration increased by 20\% during our measurement period \textcolor{black}{of 3 months}. This shows promise for \sysname becoming a useful tool in the future. 

\begin{figure}[t]
    \centering
        \includegraphics[width=0.7\columnwidth]{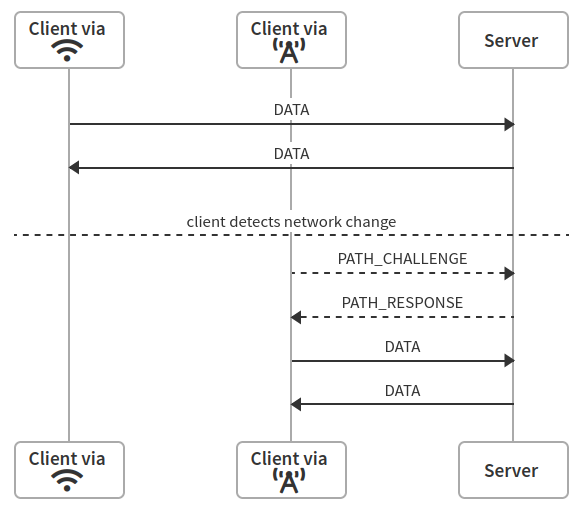}
        \caption{An illustration of QUIC connection migration. Before the server can receive data from the client on the new network path, it must be validated. The server can cache recent path validations, preventing the need to perform them every time a network migration occurs.}
        \label{fig:conn-mig}
\end{figure} 

\begin{figure*}[t]
    \centering
        \begin{subfigure}[t]{0.275\textwidth}
        \centering
        \includegraphics[width=0.85\linewidth]{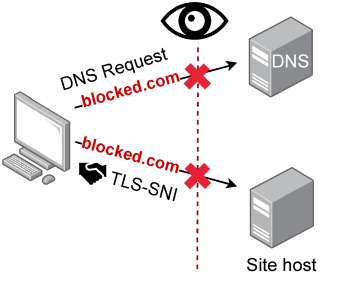}
        \caption{Adversary model}
        \label{fig:censorship}
        \end{subfigure}
        \begin{subfigure}[t]{0.4\textwidth}
        \centering
        \includegraphics[width=0.85\linewidth]{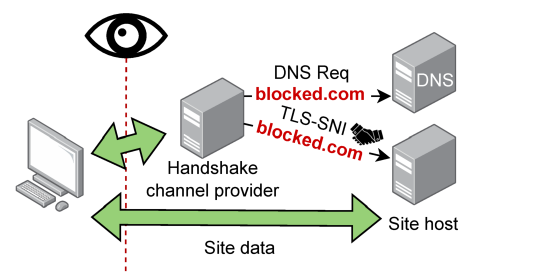}
        \caption{\sysname design}
        \label{fig:architecture}
        \end{subfigure}
        \begin{subfigure}[t]{0.265\textwidth}
        \includegraphics[width=0.85\linewidth]{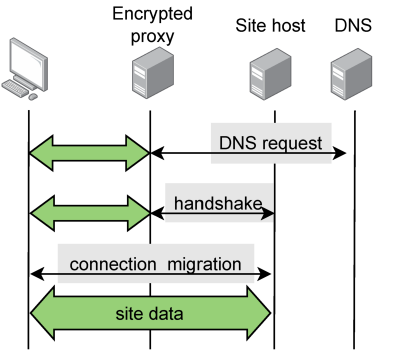}
        \caption{\sysname network requests}
        \label{fig:implementation}
        \end{subfigure}

        \caption{This figure demonstrates our adversary model and how \sysname can be leveraged to circumvent censorship. (a) demonstrates an adversary capable of monitoring, blocking or disrupting client traffic based on plaintext sensitive fields that may be associated with HTTPS requests. (b) illustrates the architecture of \sysname under this adversary model. Finally, (c) demonstrates, at a high level, the full set of network requests performed by \sysname.}
\end{figure*}

\section{Background and related work}


QUIC is a transport layer protocol based on UDP and supports multiplexing of application-layer data streams \cite{iyengar2021rfc}. QUIC 
 was developed to improve the performance of web applications compared to TCP+TLS and is the basis for HTTP/3. 
The meteoric rise of QUIC is likely to continue as HTTP/3 has been standardized as an RFC~\cite{bishop2022rfc}.  All major browsers and around 22\% of top 1 million websites already support QUIC based on our recent measurement~(\secref{sec:quicstat}).

QUIC provides a variety of network performance features.
For instance, by rolling together the QUIC and TLS handshake it eliminates a handshake round-trip~\cite{rfc9001}.
A critical performance feature for our work is \emph{connection migration}, which enables QUIC connections to persist across multiple network-layer sessions.


\paragraph{Connection migration.} QUIC utilizes a set of connection identifiers rather than the IP address and port tuple in order to uniquely identify connections. 
The decoupling of QUIC connections from IP addresses and ports allows connections to be maintained even as clients move between different networks.
This capability is referred to as \emph{connection migration}~\cite[\S 9]{bishop2022rfc}. 
When an endpoint detects a network change, it performs a round-trip \emph{path validation} to ensure that the peer is still reachable before resuming the connection, as demonstrated in Figure~\ref{fig:conn-mig}~\cite[\S 8.2]{bishop2022rfc}. 
Connection migration can only occur after a session has been fully established between a client and server through a QUIC-TLS handshake. 
QUIC connection migration enables massive performance improvements for mobile users, as connections persist even when devices move across networks such as between a mobile network and a local WiFi network. 
\sysname leverages this performance feature to circumvent censorship with minimal latency overhead.

\paragraph{QUIC-TLS handshake.} 
QUIC is designed as a secure-by-default protocol that mandates encryption of data in transit. 
To achieve this, QUIC is integrated with TLS encryption.
Notably, QUIC-TLS encrypts the \texttt{Initial} packets with a secret derived from a public salt~\cite[\S 5.2]{rfc9001}. 
Although QUIC \texttt{Initial} packet encryption does not ensure confidentiality—since the keys can be derived by anyone observing the connection—it still complicates SNI-based censorship by DPI middleboxes as it requires additional computational resources for decryption and more effort to track UDP flows.

\subsection{Censorship of QUIC traffic}
Network-level adversaries can analyze QUIC-TLS handshake packets and censor connections based on the TLS SNI field, 
making HTTP/3 connections vulnerable to censorship.
Specifically, a censor can first compute the secret used to encrypt the QUIC client's \texttt{Initial} packets with the client's destination ID and the public salt~\cite[\S 5.2]{rfc9001}.
It then decrypts the client's \texttt{Initial} packet
and extracts the SNI field in its TLS \texttt{ClientHello} message.
If the SNI field matches the censor's blocklist, the censor can then block the ongoing QUIC connection.

The development and adoption of the QUIC protocol created a temporary gap in QUIC censorship as censors needed time to develop DPI software and devices capable of inspecting and blocking QUIC traffic.
For instance in 2023, following the Turkish government's decision to block a social media website due to criticism over its handling of the T{\"u}rkiye-Syria earthquake fallout, the website's developers reported that users could circumvent TLS-SNI-based censorship by forcing a QUIC connection to their server~\cite{eksisozluk}.

However with the rise of QUIC traffic, censorship regimes continue to devise methods to detect and block QUIC traffic. 
%
Researchers reported multiple instances of QUIC blocking around the world, including China~\cite{China-QUIC-Censorship}, Uganda~\cite{elmenhorst2022blog},
Iran~\cite{elmenhorst2021web}, and Russia~\cite{elmenhorst2022blog}.
Early attempts of QUIC censorship included blocking QUIC traffic in general.
For example, in 2022 Xue et al. reported that the Russian TSPU censored QUIC traffic by identifying packets that have certain fingerprints, then dropping all packets in that flow~\cite{xue2022tspu}.
This strategy aims to censor all QUIC traffic and cannot selectively censor traffic to certain websites.
As QUIC traffic is expected to increase, broad QUIC blocking may yield increasing amounts of collateral damage.

As detailed in~\cite{China-QUIC-Censorship}, since April 2024, the GFW started censoring QUIC traffic by first decrypting the QUIC client's \texttt{Initial} packets then inspecting if the SNI field in the TLS \texttt{ClientHello} message matches the blocklist. 

\subsection{QUIC connection migration for privacy}

\subsubsection{CoMPS} \cam{CoMPS is a connection-migration traffic splitting framework proposed by Wang et al. aiming to improve robustness against website fingerprinting~\cite{wang2022leveraging}.
CoMPS uses a path scheduler that splits traffic
across multiple network paths to limit the amount of traffic an adversary can observe, under the assumption that the adversary can only observe packets on a single path.
The client uses connection migration to route traffic across the different paths.
This work was also the first to present a high-level sketch of \sysname, suggesting that by sending handshake packets through a VPN path, CoMPS can be used to circumvent SNI-based censorship.}
However, censorship circumvention is only mentioned briefly as a potential use case of CoMPS, and is not implemented or evaluated for deployment, feasibility, or performance.
Thus, we explore the following open questions: Would connection-migration based circumvention be effective in practice? What would the performance benefits be compared to tunneling all traffic using the handshake channel?

\subsubsection{MIMIQ} \cam{MIMIQ 
uses connection migration in QUIC to frequently change client IP address within a trusted network to thwart user tracking and certain types of traffic analysis attacks~\cite{govil2020mimiq}.
MIMIQ requires cooperation from the client network, as it needs the network to set up modified DHCP and an edge switch.
Thus it is not suitable for our censorship circumvention scenario where the client network itself may be adversarial.} 


\subsection{Related work}

\subsubsection{TLS session resumption}
TLS session resumption has been proposed as a way to circumvent SNI censorship.
Introduced in TLS 1.2, session resumption reduces the need for clients to conduct handshakes for each TLS connection.
When a TLS session is first established, the server sends a unique ticket to the client which can be used by the client to resume a TLS session with the server.
MultiFlow and REDACT propose using session resumption to enable decoy routing~\cite{manfredi2018multiflow, devraj2021redact}.
More recently, BlindTLS proposes establishing a connection to a censored domain through a VPN proxy and resuming the session in plain sight of the censor with a different SNI using TLS session resumption~\cite{Satija2021a}.
\textcolor{black}{However, BlindTLS focuses on session resumption in TLS 1.2, and would not work with TLS 1.3 as TLS 1.3 requires that the SNI sent in the resumption handshake matches the SNI associated with the session~\cite{rfc8446}.} In contrast with BlindTLS, \sysname is compatible with TLS 1.3 and is designed to be independent of TLS version. 

Our work is the first to thoroughly investigate leveraging various properties of QUIC as a transport protocol to circumvent handshake censorship, rather than being dependent on application-layer features of TLS.
Prior SNI censorship circumvention literature has focused on the traditional TCP setting as most censorship in the wild has been observed in this setting~\cite{Chai2019a, Hoang2024a}.

\subsubsection{Encrypted ClientHello}
\cam{Encrypted ClientHello~(ECH) is an extension to TLS 1.3 that encrypts the ClientHello message, including the server name, to protect user privacy~\cite{ech-25}.
ECH could be used to circumvent censorship based on SNI blocklists since the SNI is hidden from observers.
However prior research has found that ECH support in servers is limited and that censors in Russia, China, and Iran currently censor ECH traffic, undermining ECH's practicality in censorship circumvention~\cite{niere2025encrypted}.}

\subsubsection{Measurement of QUIC connection migration support}

Directly relevant to our work is the availability and popularity of connection migration.
In 2024, Buchet and Pelsser investigated QUIC connection migration support on the web~\cite{buchet2024analysisquicconnectionmigration}.
This work tests connection migration by sending a packet with a new connection ID to the server, but does not initiate or test if they are able to successfully load resources after migrating connections.
We found that such a method was incomplete in measuring the practical success rate of migrating connections after switching to a different port or IP address.
For instance, QUIC servers generally seem to support port-based connection migration (e.g. migrating a QUIC connection onto a new UDP port) differently from IP-based connection migration (migrating a QUIC connection onto a new IP address).
Our work leverages \sysname to provide the most complete Internet-wide connection migration support measurement to date.
We additionally use data from these results to provide recommendations to standards bodies for ways to improve the usefulness of connection migration for all QUIC clients.


\subsubsection{Other censorship circumvention systems} \label{sec:bk:censor-attack}
There is a long line of research on censorship circumvention 
that assume more powerful adversaries compared to \sysname's threat model.
These works use complex traffic shaping, traffic mimicry, and other obfuscation techniques for tunneling encrypted traffic, often against adversaries that can use higher-cost techniques (e.g. powerful machine learning classifiers) to identify blocked content from the metadata of encrypted traffic flows~\cite{mohajeri2012skypemorph,houmansadr2013freewave, rosen2021balboa,figueira2022stegozoa,jia2023voiceover}.
While we briefly discuss more powerful adversaries in \secref{sec-analysis}, we note that the focus of this work is on realistic adversaries that have more limited resources.
\sysname's focus on a ``lightweight'' censor is similar to 
Geneva, domain fronting, domain shadowing, and Snowflake, many of which are deployed in practice and are used by millions of users to circumvent network censorship~\cite{bock2019geneva,fifield2015blocking,domain-shadowing,snowflake}.


\subsubsection{Real-world censors prefer simple and efficient detection methods} \label{sec:bk:censor}
As the GFW is often a first-mover in the global censorship ecosystem, we soon expect other censorship regimes to follow suit in enacting QUIC SNI censorship~\cite{China-QUIC-Censorship}.
The goal of this work is to stay ahead of the censors: what might widespread, global QUIC SNI censorship look like in practice?
Empirical measurements of real-world censorship machines reveal that censors prefer relatively simple and efficient detection mechanisms over comprehensive but complex or expensive detection mechanisms~\cite{cat-and-mouse}.
For example, Wu et al. discovered that the GFW only inspects the first TCP payload sent by the clients when detecting if a TCP connection is fully encrypted~\cite{Wu2023a}.
This is consistent with the idea that the handshake packet, even when encrypted, is the most information-rich portion of the connection that are used by censors to make efficient blocking decisions.
Similarly, Zohaib et al. discovered that the GFW assumes the first UDP payload is a complete QUIC client \texttt{Initial} packet when conducting QUIC SNI-based censorship~\cite{China-QUIC-Censorship}.
\cam{Based on these observations we focus on a censor that performs stateless, lightweight censorship.}

\cam{We acknowledge that in the future more invasive and powerful censors that perform stateful censorship may emerge. 
But given that 
GFW is the primary QUIC-SNI censorship system that is deployed in the real world, we focus on GFW-style censors and consider stateful censors beyond the scope of this work.}
We discuss our threat model further in section \secref{sec:threatmodel}.

\section{Bringing \sysname from theory to practice}

\sysname's primary goal is to circumvent QUIC SNI censorship while minimizing overall latency overhead and avoiding modifications to server-side software or the QUIC protocol. In this section, we discuss how we bring \sysname from theory to practice, by first demonstrating our threat model and implementation goals, and then delving into the challenges and implementation details.

\subsection{Threat model}\label{sec:threatmodel}

Our threat model follows prior work on research on censorship circumvention techniques such as domain fronting~\cite{fifield2015blocking} and Geneva~\cite{bock2019geneva}.
In our threat model, the client is located in a censored network and aims to access a censored domain hosted on a \emph{non-blocked} IP address outside the censored network.
The censor uses DPI techniques such as DNS and SNI filtering to identify and prevent such access.
We consider a practical censor as discussed in \secref{sec:bk:censor}, who employs lightweight methods to achieve real-time detection at scale.
Specifically, the censor leverages \emph{stateless} detection techniques that can be performed on a per-packet basis, and does not record all network flows to perform flow-level traffic analysis.
Anonymity is not a primary concern for the client, which aligns with the assumptions in previous works (e.g., MassBrowser~\cite{nasr2020massbrowser}).
In practice, public VPNs or proxies are commonly used by users in censored countries to bypass censorship even though these tools do not guarantee anonymity.
User surveys also suggest that users in censored regimes often prioritize content access and internet speed over security or anonymity~\cite{callanan2011leaping, xue2024bridging}.




\subsubsection{Client model and the \channelname} \label{design:client}

The client wants to access censored domains with low performance overhead.
We assume that the client \textcolor{black}{can already access a secure, blocking resistant but potentially high-latency \channelname (e.g. a DNS tunnel or public VPN) for every connection.}
A large amount of censorship resilience literature is focused on the development of such high-security, low-bandwidth channels, such as Tor bridges/pluggable transports, the \emph{rendezvous channel} used in Tor's Snowflake~\cite{snowflake}, \emph{signaling channels} for Tor bridge distribution~\cite{Vines2024c}.
The client can employ any of these synchronous, high-security channels as a \channelname.
\cam{Censorship circumvention tools like Hysteria~\cite{hysteria}, V2Ray~\cite{v2ray}, Xray~\cite{xray}, Sing-box~\cite{sing-box}, and Tor pluggable transports such as Meek (domain fronting)~\cite{meek} are also potential options for handshake channels.}

However it is not desirable for the client to rely on this channel for all traffic.
The \channelname is resource-constrained (shared with other users), and thus may charge or impose bandwidth or data constraints on individual users. 
For example, the free version of Lantern has a monthly data limit of 500MB~\cite{lanternlimit}.

We also assume that the censor is unable to break the security guarantees of QUIC or the \channelname and will not attack the availability of certain classes of web traffic (e.g. blocking all QUIC traffic) to avoid collateral damage.
Further discussion on blocking all or certain types of QUIC traffic continues on \secref{sec-analysis}. 
%
Figure \ref{fig:censorship} illustrates a censored network representing our threat model.

\subsection{Implementation} \label{sec:implementation}


Figure~\ref{fig:architecture} depicts the high-level architecture of \sysname.
The client first completes a QUIC-TLS handshake and establishes a QUIC session with the server through a secure \channelname.
The client then immediately switches to the native network path, allowing the rest of the packets to be sent directly to the server with minimal latency. 

\sysname requires setting up a \channelname. The \channelname can be any secure channel with blocking resistance (e.g., VPN). For our prototype, we used a WireGuard channel. \textcolor{black}{We did not use Tor even though it 
provides much higher security and anonymity guarantees, as Tor currently does not support UDP tunneling.} Furthermore, using WireGuard proxies under our control enables a more controlled experiment where we can vary proxy location or throughput. Many users in censored domains actively use such custom proxies for censorship circumvention~\cite{shadowsocks-python, shadowsocks-rust, v2ray, xray, sing-box, hysteria}.
We note that our implementation of \sysname is agnostic to the particular type of \channelname so long as the channel provides a virtual network interface.


\subsubsection{Implementation goal}
Our primary goal in implementing \sysname is to develop an easy-to-implement, application-agnostic solution.
We want \sysname to be compatible with the vast technological environments that clients and servers may be running in, e.g., requiring no modifications to client applications, upper-layer protocols, operating systems, and browsers.
Our implementation of \sysname does not necessitate any changes apart from requiring that the client and server support QUIC and connection migration.

\subsubsection{Implementation choices and challenges} \label{sec:impl-choices}
Once handshake happens through the \channelname, the challenge is to identify confirmation of the handshake and route packets accordingly (\cam{regardless of the application through which the packets are transmitted)}.
According to the QUIC standard, connection migration is expected to happen after the handshake is \textit{confirmed} at the peer~\cite{iyengar2021rfc}. 
When the server confirms the handshake, it sends a \texttt{HANDSHAKE\_DONE} frame to the client in a 1-RTT packet, but this frame is encrypted and hard to identify \cam{from outside of the application}.
If handshake confirmation is not accurately identified it can incur latency overhead as the server sends additional packets through the \channelname to \textit{complete} the handshake.

To accurately identify handshake confirmation we considered two options: (1) using eBPF~\cite{ebpf} to heuristically determine which connections had finished handshake confirmation based on unencrypted parts of payloads and (2) modifying QUIC clients to track the frame.
However, the first solution requires managing the state of multiple connections and requires privileged deployment in the client, and the second would be challenging to deploy in practice as it requires changes to the QUIC client.
Neither of these satisfied our requirements for flexibility and ease of future deployment.

As such, we chose instead to {\emph{approximate}} handshake confirmation time by simply routing handshake packets differently from data packets.
The key insight is that QUIC handshake packets use the QUIC ``long header'' format, but data packets use the QUIC ``short header'' format. The two formats are differentiated by the unencrypted first bit of the QUIC header.
Thus it is possible to differentiate handshake packets from data packets using only information exposed on the wire.


\subsubsection{Proof-of-concept implementation description.} 
Our proof-of-concept version of \sysname uses \texttt{iptables} firewall rules to route select packets through the WireGuard interface.
Our rules route DNS packets (UDP packets headed to port 53), TCP packets, and QUIC long header packets (UDP packets headed to port 443 whose payloads begin with 1) through the WireGuard interface.
This ensures that all packets containing server name are transmitted securely. In addition, \sysname does not have any requirements for the client besides having access to a WireGuard proxy (or other \channelname), and incurs no significant latency overhead on the client.
\cam{If the server does not support QUIC or connection migration, the client would fall back to TCP which is fully transmitted through the \channelname. The user experiences no significant failure.}

The code for our implementation and evaluations are made available at \cam{\url{https://github.com/inspire-group/QUICstep}}. 

\section{Evaluation}






In this section, we evaluate \sysname's ability to circumvent real-world censorship as well as its performance.
\subsection{Research questions and overview}
We give an overview of the evaluations designed to answer each of our research questions (as outlined in \secref{sec:intro}). 

\subsubsection{\sysname-compatibility: measuring compatible websites in the wild} We identify the current state of QUIC and connection migration support among popular websites by performing HTTP/3 GET requests with \sysname~(\secref{sec:quicstat}).
We find that 22\% \textcolor{black}{($\sim$220\,K)} of top 1\,M domains support QUIC and 12.8\% \textcolor{black}{($\sim$28\,K)} of QUIC supporting domains are compatible with \sysname. Our findings \textcolor{black}{additionally} suggest that the role of service providers is critical in support for connection migration.

\subsubsection{Effectiveness: practical censorship circumvention with \sysname}
We evaluate \sysname's ability to circumvent real-world censors that perform SNI-based censorship on TLS and QUIC traffic~(\secref{sec:real-world-censor}).
Since \sysname transmits handshake packets through a 
\channelname,  
it becomes much more difficult for censors to judge whether to block a particular connection or not.
We find that \sysname indeed successfully circumvents SNI-based censorship in the wild, including the recently implemented QUIC SNI censorship by the GFW.

\subsubsection{Performance evaluation} We \textcolor{black}{provide a quantitative analysis of} the latency of \sysname to scenarios where all traffic is sent through the native channel or the \channelname \textcolor{black}{under different settings}~(\secref{sec:quicperf}).
QUIC connection migration ensures that path switching between the direct and tunneled paths only introduces one RTT of latency (i.e., path validation), and does not disrupt the ongoing session between the client and the server.
Compared to a native QUIC connection, \sysname does incur some additional latency due to conducting the handshake over the \channelname and path validation during path switching.
However, we find that this latency overhead is amortized over the entire request and \sysname provides up to 84\% reduction in page load time compared to \cam{using the \channelname for all traffic}.

We also show that \sysname reduces load on the \channelname provider by a median of 93\% over 100 different websites~(\secref{sec:eval:proxy-load}). High-latency channels incur hosting costs, such as for rendezvous hosts, Tor bridge volunteers, or for resilient VPN providers. Bandwidth is either explicitly limited due to the mode of transport, otherwise capped by providers, or simply lowered in practice due to the channel being often overloaded. By opportunistically migrating connections to the devices' native network, \sysname minimizes bandwidth usage and reduces load for the high-latency channel.




\subsection{\sysname support: measuring compatible websites in the wild}\label{sec:quicstat}
To identify the scale of \sysname's impact, we aim to measure the number of websites (domains) that support QUIC, are fully compatible with \sysname, or only partially support connection migration.
Fully compatible websites are ready to use with \sysname, while QUIC-supporting websites that do not or partially support connection migration will benefit from \sysname in the future as QUIC libraries continue to mature.

\subsubsection{Experiment setup and methodology}
We begin with identifying QUIC-supporting websites in the wild.
To identify server-side support for QUIC, we sent HTTP/3 GET requests over QUIC to Tranco top 1\,M domains~\cite{LePochat2019}, denoting support for QUIC if the client successfully connects to the server.
We used the Chromium \texttt{quic\_client} to perform these tests~\cite{quiche}.
We then aim to find \sysname-compatible websites among QUIC-support websites.
To measure \sysname compatibility we conducted HTTP/3 GET requests with \sysname enabled and denoted success when the fetch succeeded without error.
\cam{We examined packet captures during the connection for a sample of successful \sysname fetches and verified that the connection was functioning after migration.}
We repeated requests for both the parent and \texttt{www} subdomain.
We denoted success if either of the two succeeded.
We excluded 404 Error pages served over QUIC, but included redirect pages.
The client and \qsproxy for \sysname were located in North Virginia and Ohio, respectively. 


\subsubsection{A notable portion of QUIC websites are \sysname-compatible} \label{sec:quicstep-compatible}
As of October 25, 2024, we found 219,729 websites (22\%) among top 1\,M that support QUIC.
Out of these QUIC-supporting websites, 28,104 (12.8\%) were compatible with \sysname.
While QUIC support is yet far from universal, we expect that in the future QUIC will be both more widespread (more entities supporting QUIC) and complete (QUIC implementations fully adhering to standards).

We further analyzed the associated network providers of these compatible domains.
\tableref{table:quic-provider} shows top 10 QUIC providers within the top 1\,M domains and \sysname-compatibility in each provider.
As observed, 74.6\% of QUIC-supporting domains are using Cloudflare; however, only a small fraction (0.2\%) of Cloudflare's QUIC-supporting domains were compatible with \sysname.
Upon further investigation, we found that this low compatibility rate stems from a specific limitation within the Cloudflare CDN: it does not yet fully support connection migration. Since connection migration support is expected as per the QUIC RFC~\cite{iyengar2021rfc}, this demonstrates that even broadly-used implementations of QUIC are still lagging behind the expectations of the full specification.
We also tested Cloudflare's open-source QUIC implementation, quiche, and confirmed that it currently lacks proper support for connection migration. 

Hetzner has a particularly high support of connection migration among their QUIC-supporting domains.
Most of the connection migration supporting domains in Hetzner (likely using Hetzner's hosting service) use the Litespeed server, which provides connection migration capability.
This shows major cloud/hosting providers play a key role in the rollout of connection migration; a simple service update may significantly increase the number of websites able to benefit from connection migration.

We argue that some domain owners may be unaware they are missing out on QUIC's performance benefits due to the incompatibility of their websites' dependencies (CDN, load balancer, etc.) with standard QUIC. 
Some dependencies do not fully support all critical QUIC features defined in the RFC standards such as connection migration despite announcing QUIC support.

Another case is AWS CloudFront. CloudFront claims to support connection migration, but in practice support was often inconsistent.
We will discuss more of this in \secref{sec:eval:partial}.


\begin{table}[t]
\centering
\begin{tabular}{|c|c|c|}
\hline
\textbf{Provider} & \textbf{\# QUIC domain} & \textbf{\begin{tabular}[c]{@{}c@{}}\# of \sysname-compatible\\domains \end{tabular}} \\ \hline
Cloudflare & 163900 (16.4\%) & 335 (0.2\%) \\ \hline
Google & 7191 (0.72\%) & 748 (10.4\%) \\ \hline
Amazon & 7067 (0.71\%)  & 3858 (54.6\%) \\ \hline
Hostinger & 4738 (0.48\%) & 2770 (58.5\%) \\ \hline
Fastly & 4140 (0.41\%) & 2114 (51.1\%) \\ \hline
Hetzner & 2187 (0.22\%) & 1802 (82.4\%) \\ \hline
Automattic & 1675 (0.17\%) & 9 (0.5\%) \\ \hline
Wix & 1084 (0.11\%) & 0 (0.0\%) \\ \hline
OVH & 1068 (0.11\%) & 710 (66.5\%) \\ \hline
Bigcommerce & 896 (0.09\%) & 1 (0.1\%) \\ \hline
\end{tabular}
\caption{Top 10 QUIC providers and the proportion of \sysname compatibility in each.}
\label{table:quic-provider}
\end{table}




\subsubsection{The number of websites with partial support for connection migration is growing}\label{sec:eval:partial}
In our measurements, we found websites that do not support IP address migration but support port migration.
This is likely because QUIC libraries that web servers or their dependencies use fail to properly implement the connection migration feature.
We anticipate these partial supporters could become compatible with \sysname in the future as QUIC libraries continue to mature.

\begin{figure}[t]
 \centering
     \includegraphics[width=\columnwidth]{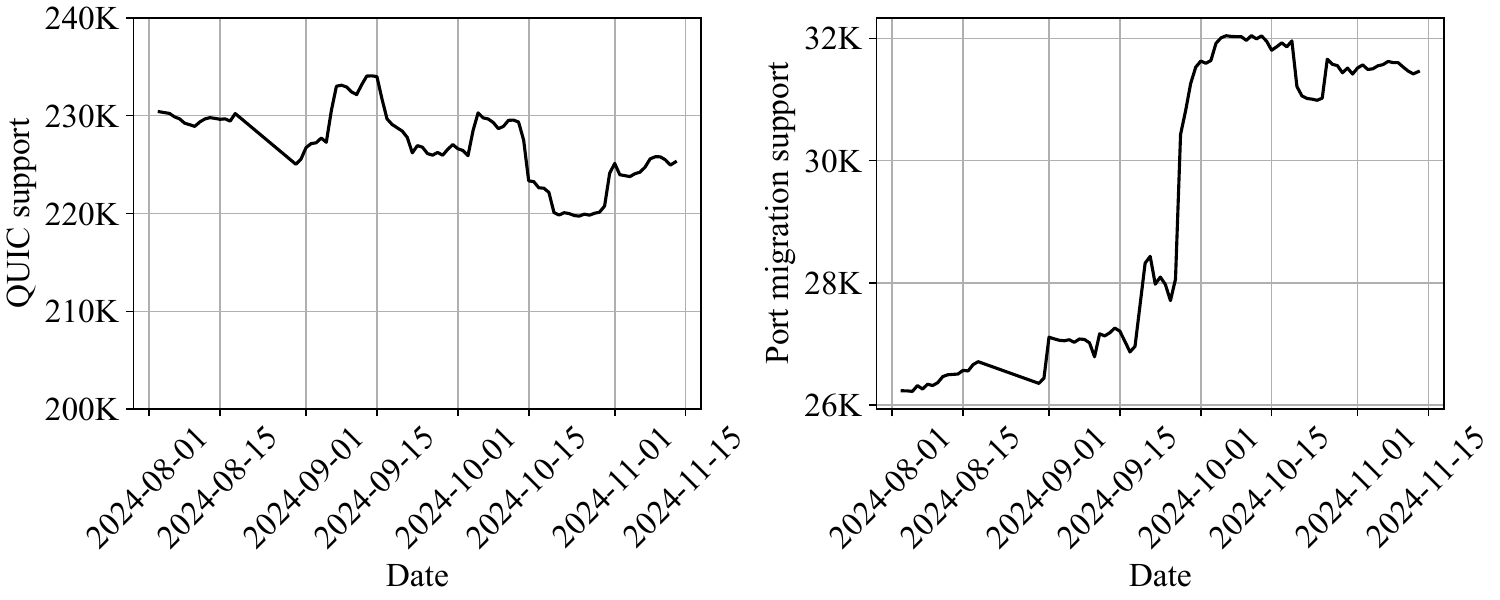}
    \caption{Number of websites that support (a) QUIC and (b) port migration from daily Tranco top 1M websites from August 3, 2024 to November 13, 2024. Support for port migration increased sharply around late September, 2024.}
    \label{fig:trend}
\end{figure}

We conducted a long-term measurement to track these \emph{potential} \sysname-compatible websites.
To measure port migration support, we leverage the ephemeral port migration option provided in the \texttt{quic\_client}.
This option performs an HTTP/3 GET request and then migrates to an ephemeral port to perform a second HTTP/3 GET request on the same connection.
We note that some of those port migration supporters may already support IP address migration, which makes them fully \sysname-compatible; nevertheless, we count them as potential.
We performed daily scans on top 1\,M domains from August 3, 2024 to November 13, 2024.
\figref{fig:trend} shows that while the number of QUIC-supporting websites has been fluctuating, the number of potential \sysname-compatible websites has increased significantly.
When we began measurements on August 3, 2024, 26,234 domains supported port migration.
On September 26, 28,060 websites supported port migration; by September 29 it increased to 31,262 and the number has stayed above 31,000 since.

The increase in potential \sysname-compatible websites could be attributed to QUIC library upgrades.
For example, Varnish CDN did not support connection migration when we began the measurement, but our recent check confirms that it now fully supports port migration and some of its servers support IP address migration. 

There are some special cases that support only IP address migration and not port migration.
We exclude this case from measurement because nearly all of these domains are associated with the CloudFront CDN.
Additionally, we found that different servers (distinguished by IP addresses) that host \texttt{amazon} domains have varying responses to IP address migration.
Even for the same domain, some servers support IP address migration but some do not.
We suspect there were deployment issues that caused the servers to use different versions of QUIC implementations or configurations (e.g., we have received ``connection migration disabled by server'' error messages \textcolor{black}{from some servers but not others}).
We expect that robust and consistent migration support will extend to all servers in the near future.

\subsubsection{The road to connection migration}



Our measurements indicate that while connection migration adoption is on the rise, two major practical challenges could impede further deployment and usage: (1) QUIC implementations are not compliant with the RFC standard, and do not or only partially support connection migration. (2) The complex dependencies of modern websites demand robust support for connection migration at critical infrastructure points.
Tackling these challenges requires cooperation between different stakeholders such as QUIC developers, application providers, and service providers.
Particularly, due to the centralization of the Internet, service providers play a critical role in the process; their willingness to adopt updated implementations and configure the infrastructure to support connection migration will directly impact the feature's success.
For example, if Cloudflare properly supports connection migration, \textbf{87.2\%} of QUIC-supporting domains would be able to leverage connection migration. 

We believe that a crucial step in advancing connection migration rollout is to develop a reliable tool for testing connection migration in both QUIC library development and QUIC-supporting service deployment.
We see \sysname as a stepping stone.
As discussed above, connection migration involves complex cases that have often been overlooked and can only be discovered during actual migration events e.g., supporting port but not IP address migration.
\sysname can be viewed as a system that introduces artificial mobility events into the network to activate the migration features of native QUIC.
With \sysname, there is no need to \emph{infer} a server’s support for connection migration using indirect information, as done in previous work~\cite{buchet2024analysisquicconnectionmigration}. 
Instead, \sysname obtains a \emph{definitive} confirmation of migration support based on whether the migration has actually succeeded.

\subsection{Effectiveness: practical censorship circumvention with \sysname}
\label{sec:real-world-censor}
\sjl{We test the capability of \sysname to circumvent real-world SNI-based censors, confirming that \sysname is capable of accessing blocked websites.}



\subsubsection{Real-world TCP+TLS SNI censorship} \label{sec:real-tcp}
Our first evaluation was against our local [anonymized institution] Palo Alto Networks firewall that performs TLS SNI-based censorship.
We accessed a domain under our control that was blocked by the firewall via SNI censorship.
We did not control the firewall, and the domain was added to the firewall by Palo Alto's automatic system due to the domain name being relevant to cryptocurrency.

Our tests verified that the domain was blocked via SNI-based censorship.
DNS requests to this domain returned the correct IP address, TCP SYN -> SYN+ACK handshake with the IP address completed successfully, and directly connecting to the IP address without domain name also succeeded.
However, we observed a middlebox inserting an RST packet into the connection immediately after the client sent a TLS \texttt{ClientHello} with this domain name in the SNI field.

We successfully established a QUIC session with this domain via \sysname.
We note that connecting to the web server with a native QUIC connection also succeeded, implying that SNI censorship performed by this firewall did not perform SNI-based blocking in QUIC traffic.
Nevertheless, this does not diminish \sysname's success as our findings confirm that \sysname's applicability includes conventional TCP SNI censorship.

\subsubsection{Real-world QUIC SNI censorship} \label{sec:real-world-quic}
We take particular interest in the case of GFW, one of the leading censorship systems, has recently begun selective censorship of QUIC traffic using QUIC SNI since April 2024~\cite{China-QUIC-Censorship}.
We tested \sysname against real-world QUIC SNI censorship in China and found that \sysname can effectively bypass QUIC SNI censorship.
Websites that could not be reached with the native connection were reliably accessible with \sysname. 

We used an Alibaba VM in mainland China as the client and ran the  \texttt{quic\_client} to fetch websites (with SNI specified).
We consulted the authors of \cite{China-QUIC-Censorship} to understand the nature and extent of QUIC SNI censorship in China and obtained a list of QUIC SNI censored domains.
During the time span of our experiments, we varied the locations of \qsproxy across different AWS regions to avoid leaving consistent traces. 

Our first experiment used a QUIC server under our control, running on an AWS EC2 instance located in North Virginia.
We first set up this server with both a non-censored hostname \textcolor{black}{and a censored hostname (\texttt{youtube.com}). 
Both the native client and the \sysname client were able to access the server with the QUIC SNI set as the non-censored hostname}, over multiple repeated trials.
With \texttt{youtube.com} as the QUIC SNI hostname, the native QUIC client was blocked after several repeated fetches.
The blocking does not happen immediately because the censor inspects traffic in a probabilistic manner and blocks the connection to the destination after it detects unwanted traffic.
However, the \sysname client successfully accessed the server throughout 50 consecutive fetches.


We also identified real-world domains that can be unblocked with \sysname as discussed in \S\ref{sec:quic-censored-websites} and tested them.
We used a small sample of real-world websites for testing the feasibility of \sysname.
We tested 7 subdomains of \texttt{tiktokcdn.com} that were QUIC SNI blocked, but accessible with \sysname.
We repeatedly found that while the native connection failed after several fetches, \sysname consistently succeeded in accessing the domains. 

\subsubsection{One-third of websites currently censored by QUIC SNI in China could potentially benefit from using \sysname} \label{sec:quic-censored-websites}
We  measured QUIC migration support of the websites censored by QUIC SNI in China.
We find that connection migration support is high across websites that are censored under particular regimes. 
A concurrent work \cite{China-QUIC-Censorship} tested the full Tranco list ($\sim$7\,M websites) obtained on October 2, 2024\footnote{Domain list available at https://tranco-list.eu/list/664NX} and found 28,458 domain names were on the GFW's QUIC SNI blocklist: If a QUIC \texttt{Initial} packet contains any of these domains as the SNI, the connection will be dropped.
However, this serves more as a preventive measure since many of these domains do not support QUIC. 
We found that among these websites 2,404~\textbf{(8.45\%)} support QUIC and among them 828~\textbf{(34.4\%)} are compatible with \sysname.
\sysname can unblock these websites if the client can identify some unblocked IP address associated with the domains.

We additionally tested \sysname compatibility with websites that were censored by TCP SNI, as these websites could be gradually added to the GFW's QUIC SNI blocklist in the future. 
Using the methodology from~\cite{Chai2019a} with a test list of 65,153,600 domains\footnote{A combination of any domain that ever appeared in one of Alexa Top 1M, Tranco 1M, Cisco Umbrella 1M for at least one day between Jun 23, 2021 and Jun 23, 2022}, we identified 5,700,928 websites censored by TCP SNI in China.
Among these websites 3,524,808~\textbf{(61.8\%)} support QUIC and among them 3,516,979 ~\textbf{(99.8\%)} were compatible with \sysname.
This is due to a large proportion (99.2\% of QUIC supporting domains) of \texttt{blogspot.com} and \texttt{wixsite.com} subdomains within the blocklist, both of which support \sysname.

\subsection{Performance evaluation}\label{sec:quicperf}
In this section we aim to understand the factors that affect performance of \sysname \textcolor{black}{with comparison to completely relying on the proxy} under different settings.
We investigate the effect of different configurations including proxy bandwidth, client location, and proxy location on page load time and time to first byte.
We demonstrate that \sysname significantly reduces the load on a VPN proxy compared to full VPN connections, and provides greater performance gain in bandwidth-limited environments. By optimizing proxy location and DNS resolution, the performance of \sysname can be further boosted.

\begin{figure}[t]
    \centering
    \includegraphics[width=0.6\linewidth]{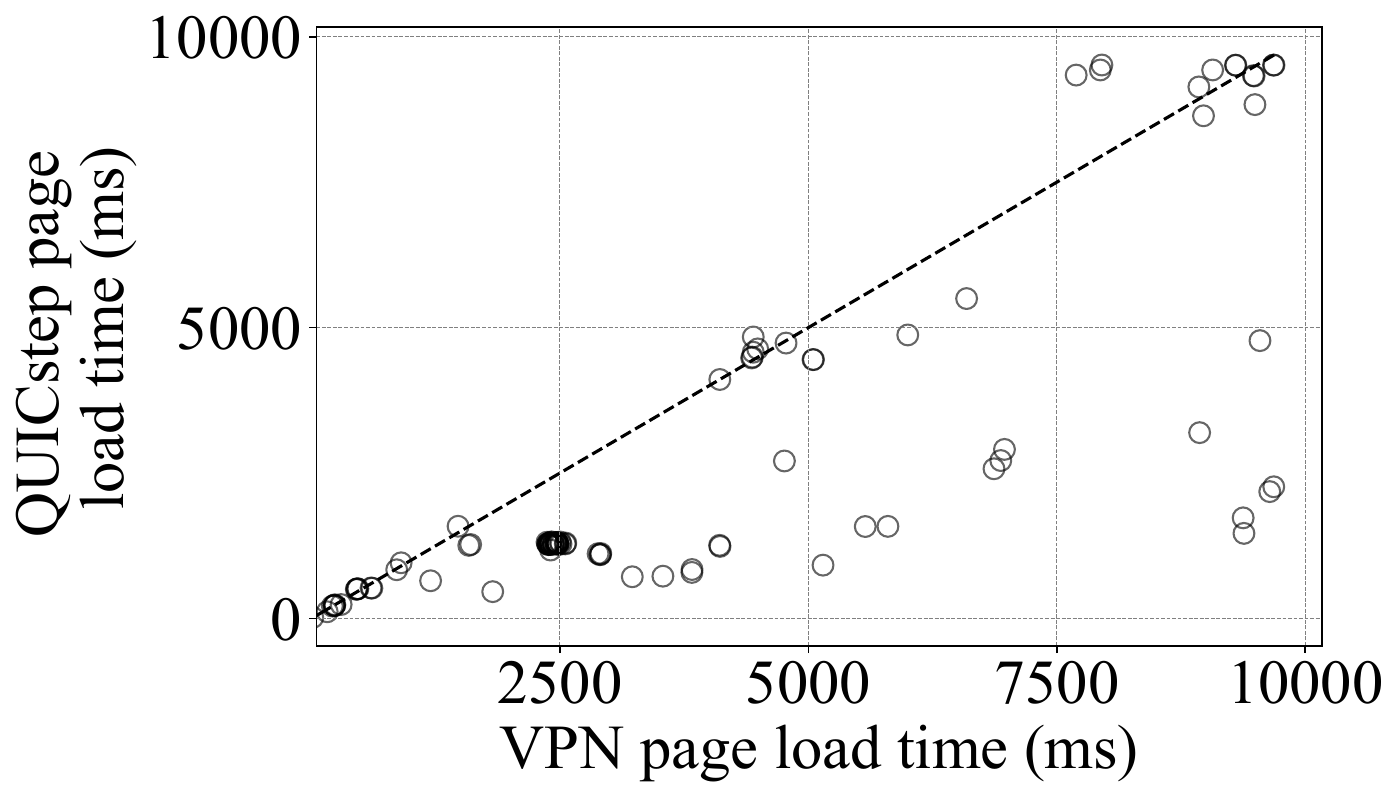}
    \caption{Page load time of \sysname and VPN connections for 100 different domains. Client in London, \qsproxy in Ohio with a maximum throughput of 5\,Mbps. \sysname generally provides shorter page load time compared to VPN.}
    \label{fig:loadtime_websites}
\end{figure}

\subsubsection{Methodology}\label{sec:eval:method}
In our default setting, our client is in London (AWS region), and the \qsproxy is in Ohio, rate-limited to a maximum throughput of 5\,Mbps.
We used Chrome controlled by Selenium to visit a website that supports connection migration through the native connection, full VPN connection (i.e., tunneling all traffic via the \qsproxy), and \sysname 100 times.
In each round, we alternated between the 3 connection types to mitigate the effects of network fluctuation.
We recorded two performance metrics through Selenium: \emph{Time to First Byte} (\texttt{responseStart-navigationStart}) and \emph{Page Load Time} (\texttt{domComplete-responseStart}).
Time to first byte (TTFB) includes the latency from the handshake, which occurs through the encrypted VPN channel.
Therefore, we expect TTFB for \sysname to be similar to that of the VPN.
However, after the handshake, \sysname fetches the website content through the native connection, so we anticipate a reduced page load time for \sysname compared to the VPN.
 
In our evaluation, we varied client location, \qsproxy location, and \channelname bandwidth to understand their influence on \sysname performance.
We conducted our measurement with three client locations (London, New Jersey, and Osaka) and seven proxy locations (Frankfurt, Ireland, Montreal, Ohio, Oregon, Seoul, and Tokyo), resulting in a total of 21 location combinations.
Besides, we set different rate limits on the \qsproxy to emulate the bandwidth-limited secure channels real-world users would have.
We tested a maximum throughput of 1\,Mbps, 5\,Mbps, 10\,Mbps, and with no rate limit. These numbers were chosen to match the scale of the throughput of popular proxy services: Tor has a median throughput of around 10\,Mbps~\cite{torthroughput}, Psiphon's free version limits throughput to 2\,Mbps~\cite{psiphonthroughput1, psiphonthroughput2}. 

Unless explicitly mentioned, our observations are generally consistent across various settings, and for clarity and simplicity, we only report on the result from the default setting (client: London, \qsproxy: Ohio, rate: 5\,Mbps).
The performance number is the median of 100 rounds of measurements. 


\begin{figure*}[t]
\centering
\includegraphics[width=\textwidth]{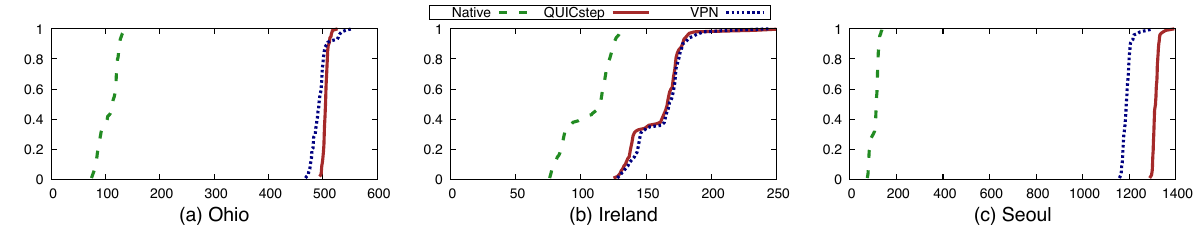}
\caption{CDF of time to first byte (in \textbf{milliseconds}) from 100 fetches of www.youtube.com with the client located in London and proxies located in Ohio, Ireland, Seoul with proxies' maximum throughput limited to 5\,Mbps. \sysname performance is comparable to the VPN connection.}
\label{fig:firstbytes}
\end{figure*}

\subsubsection{\sysname generally offers improved performance compared to full VPN} \label{sec:eval:perf}
Our measurement \textcolor{black}{quantifies the }performance gain \textcolor{black}{\sysname provides} over full VPN connections.
Figure~\ref{fig:loadtime_websites} shows the distribution of VPN and \sysname median page load time (across 20 rounds) for 100 different domains that are \sysname-compatible in the Tranco top 1\,K list from October 25, 2024\footnote{Domain list available at https://tranco-list.eu/list/N34YW}.
The majority of tested domains experience shorter page load time when using \sysname compared to VPN, with the time reduction being as great as 84\%. \textcolor{black}{As noted in \secref{sec:impl-choices}, our implementation approximates handshake completion, so with a more accurate migration we may be able to achieve even greater reduction.}

\begin{figure*}[t]
\centering
\includegraphics[width=\textwidth]{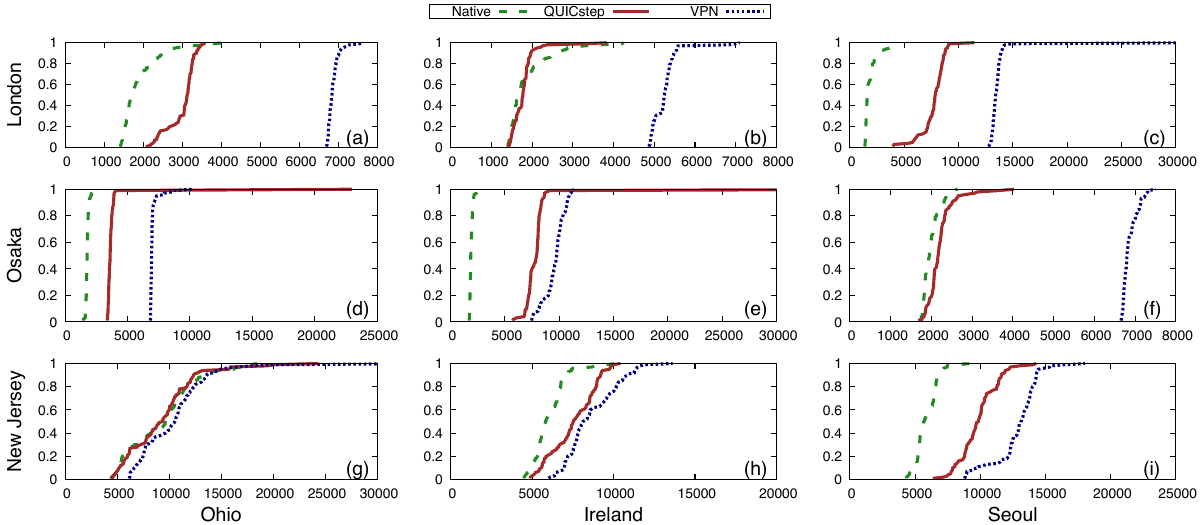}
\caption{CDF of page load time (in \textbf{milliseconds}) from 100 fetches of www.youtube.com, with  \textbf{clients} in  London, Osaka, New Jersey (Y-axis) and \textbf{\qsproxy} in Ohio, Ireland, Seoul (X-axis). The \qsproxy's maximum throughput is limited to 5\,Mbps. \sysname outperforms the VPN connection in all locations and closely follows the performance of the native connection when the client and proxy are geographically proximate such as (b), (f), (g).}
\label{fig:loadtimes}
\end{figure*}

Our measurement also confirms our hypothesis about the TTFB and page load time discussed in \secref{sec:eval:method}.
We show the TTFB and page load time of \texttt{www.youtube.com} under different client/proxy location combinations in Figure \ref{fig:firstbytes} and Figure \ref{fig:loadtimes}, respectively.
Here we chose \texttt{www.youtube.com} as our target website because it is the largest website that reliably supports connection migration.
We can clearly observe that though the TTFB of \sysname is comparable to that of full VPN, \sysname provides a significant gain in page load time compared to full VPN.

\subsubsection{\sysname significantly reduces load on proxy (\qsproxy).}
\begin{figure}[t]
    \centering
    \includegraphics[width=0.5\linewidth]{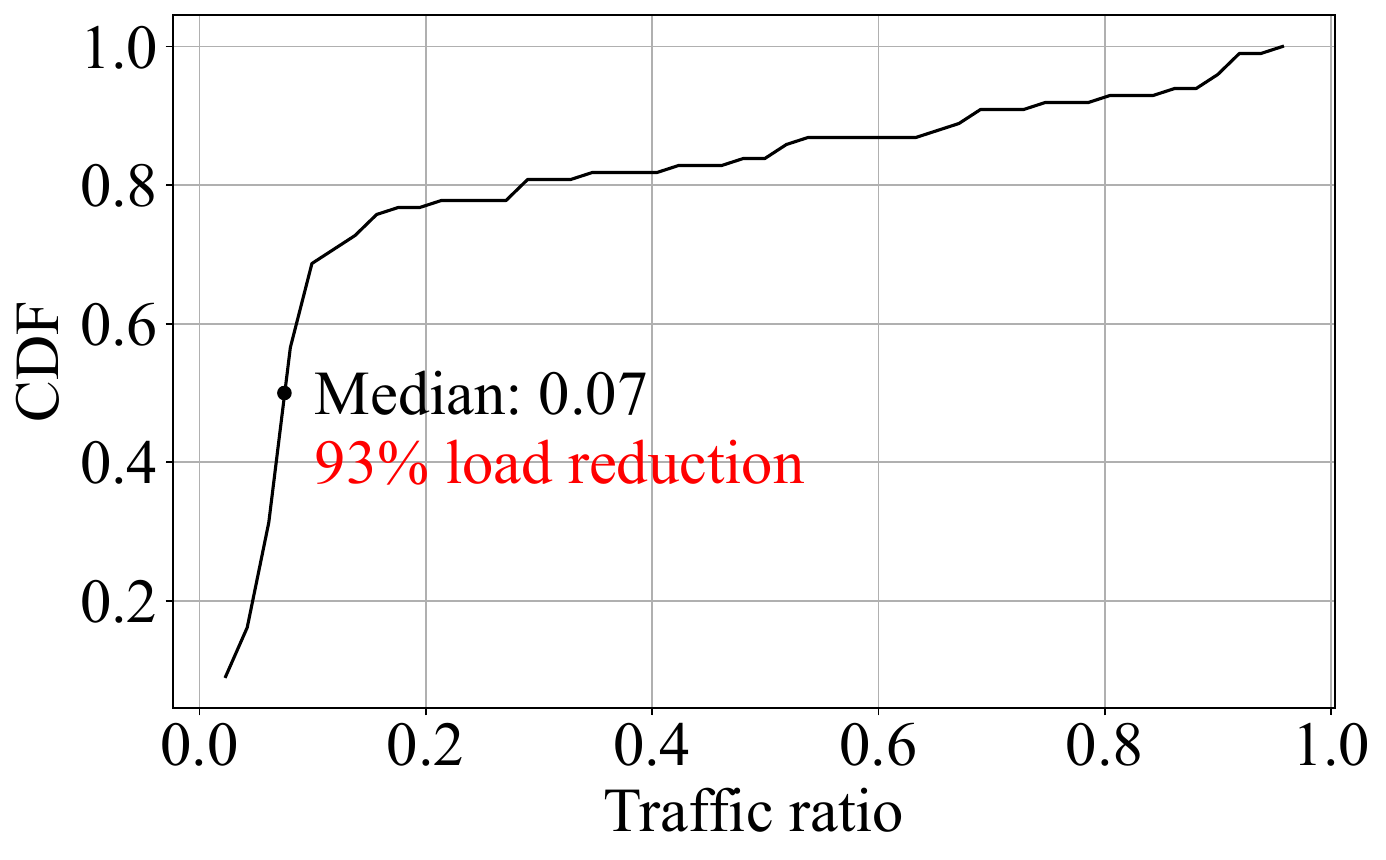}
    \caption{CDF of traffic ratio (defined in \secref{sec:eval:proxy-load}). \sysname provides a 93\% median load reduction.}
    \label{fig:proxy-load-cdf}
\end{figure}

\footnote{For brevity, `proxy' refers to \qsproxy.} \label{sec:eval:proxy-load}
The intuition behind the performance improvement is simple: \sysname can significantly reduce the traffic that needs to go through the proxy (i.e., only handshake packets in \sysname vs. full connection in conventional VPN).
\textcolor{black}{To quantitatively understand the load reduction benefit offered by \sysname, we captured the packets at the proxy during each website visit when using \sysname and VPN, and computed the traffic ratio.} \textcolor{black}{`Traffic ratio' refers to the ratio of the size of traffic through proxy in a QUICstep connection to the size of traffic through proxy in a full VPN connection, and `load reduction' refers to $1-\left(\text{traffic ratio}\right)$.}
As shown in \figref{fig:proxy-load-cdf}, \textbf{\sysname reduced the load on the VPN proxy by a median of 93\% compared to VPN}.
In the case of \texttt{www.youtube.com}, there was 3.634\,MB traffic through the proxy with the full VPN connection but only 96\,KB with the QUICstep configuration, reducing load by 97.4\%.
This load reduction helps users alleviate costs for pay-as-you-go proxy servers or enables them to maximally utilize proxy services with data limits. 

\subsubsection{\sysname can achieve performance comparable to native QUIC via optimizing proxy location} \label{sec:eval:location}

As expected, proxy (\qsproxy) location affects the performance of \sysname.
When the proxy moves farther away (geographically) from the client, \sysname adds more additional page load time to the native connection (\figref{fig:loadtimes}).
One obvious reason for this performance degradation is that the round trip time between the client and proxy increases as the proxy moves farther away.
However, another notable aspect is that changing the proxy location also changes the server's location since many websites are served through CDNs.
In \sysname (as well as VPN), the DNS request is performed through the proxy, so the client connects to a web server near the proxy; in the native connection, the client connects to a server near itself. 

We believe server location could have a greater impact on performance than proxy round-trip time, given that the majority of traffic in \sysname is sent directly to the server through the native connection.
In fact, we do observe that in certain cases, the performance of \sysname could be comparable to native QUIC (e.g., \figref{fig:loadtimes} (b) (f) (g)).
We hypothesize that in these cases our requests happened to be served by servers in the same geographical region.
This also suggests that \textbf{the client could strategically choose a proxy location that is as geographically close as possible to the client's own location that is outside the censor's regime to achieve optimal performance}, especially when connecting to websites like \texttt{www.youtube.com} that are served on CDNs consisting of geographically diverse servers.


%

\subsubsection{\sysname provides greater performance gain in the bandwidth-limited environment} \label{sec:eval:bandwidth}
Recall that real-world proxies may only provide limited bandwidth.
\channelname providers may also further downgrade services when demand is high due to resource constraints~\cite{xue2024bridging}.
To understand \sysname's performance benefit in the bandwidth-limited environment, we examined the ratio of QUICstep page load time to VPN page load time while fetching \texttt{www.youtube.com} with varying proxy maximum throughput and client/proxy locations.
A small ratio indicates a larger performance gain over VPN. 
\tableref{table:loadtime_throughput} (full table in Appendix~\tableref{table:loadtime_throughput_full}) shows the results, and we observe that \textbf{\sysname's performance gain becomes more evident as proxy bandwidth decreases}. 
Particularly, when the proxy throughput is 1\,Mbps or 5\,Mbps, \sysname always performs better than the VPN. 
When the bandwidth is 10\,Mbps, there are some cases where the VPN outperforms \sysname. 
This is likely due to AWS-based proxies being well-connected, making the proxy route potentially more efficient than the direct route between the client and server. 
Even when there is no rate limit, we were able to identify at least one proxy location where \sysname provides comparable or better page load time than the VPN connection and incurs less than 5\% latency overhead compared to the native connection. 
We observed a similar pattern for TTFB as in \secref{sec:eval:perf} and omit the results here to save space.

Overall, \sysname can provide greater performance gain when the proxy has limited throughput. 
This is particularly appealing for real-world users, given that public VPN services or volunteer proxies are often throttled. 
By reducing the amount of traffic that must pass through the rate-limited channel, QUICstep effectively mitigates potential performance bottlenecks.  


\begin{table*}[t]
\centering
\begin{tabular}{|c|c|c|c|c|c|}
\hline
 \textbf{Client} & \textbf{Proxy} & \textbf{Max throughput 1 Mbps} & \textbf{Max throughput 5 Mbps} & \textbf{Max throughput 10 Mbps} & \textbf{No rate limit} \\ 
 \hline
 \hline
  & Ireland & 0.069 & 0.338 & 0.510 & 0.996 \\
  & Frankfurt & 0.070 & 0.340 & 0.616 & 0.898 \\
  & Montreal & 0.072 & 0.463 & 0.738 & 1.052 \\
 \textbf{London} & Ohio & 0.088 & 0.458 & 0.588 & 0.996 \\
  & Oregon & 0.088 & 0.554 & 0.860 & 1.071 \\
  & Seoul & 0.128 & 0.586 & 0.633 & 0.765 \\
  & Tokyo & 0.124 & 0.682 & 0.976 & 1.553 \\
 \hline
\end{tabular}
\caption{Ratio of \sysname page load time to VPN page load time with different proxy locations and proxy rate limits. Proxies are listed in order of geographical distance from the client. The full result is in \tableref{table:loadtime_throughput_full}.}
\label{table:loadtime_throughput}
\end{table*}

\subsubsection{\sysname provides greater performance gain for large websites} \label{sec:eval:size}

Another common factor that could affect browsing performance is website size. 
However, evaluating the impact of website size on \sysname in a real-world setting is challenging, as various ``noise'' factors (e.g., dependencies on third-party servers) can affect performance estimation. 
Therefore, we performed a controlled experiment to eliminate noise. 
We set up our own QUIC server (using Google's QUICHE) and hosted files of varying sizes, and used the Chromium \texttt{quic\_client} to fetch the files under various proxy throughput limits. 
We show the ratio of file download time of \sysname over VPN in \figref{fig:size}, and more detailed numbers are in Appendix~\tableref{table:size}. As expected, we see \sysname provides greater performance gain over VPN when the file is larger.
This pattern becomes more evident when observing \sysname's overhead over the native latency (Appendix~\tableref{table:size}). 
\sysname connection does not induce additional latency after handshake, so the latency overhead stays consistent regardless of file size. 

 \begin{figure}[t]
 \centering
     \includegraphics[width=0.8\columnwidth]{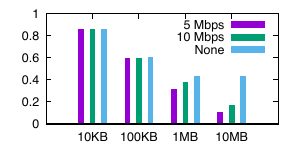}
    \caption{The ratio of file download time of \sysname over VPN under different rate limits and file sizes. A smaller ratio indicates better performance. The full result is in \tableref{table:size}.}
    \label{fig:size}
\end{figure}

We expect that for large websites, QUICstep provides greater performance gain over VPN because there is a greater proportion of traffic through the native channel. 


\subsubsection{Optimizing DNS resolution can further improve \sysname performance} 

As we discussed in \secref{sec:eval:location}, the location where DNS queries are performed has a non-negligible impact on \sysname performance, because it may consequently influence the location of the server the client will connect to.
Ideally, we would like the client to perform DNS queries by itself to ensure the selection of the most optimized server.
However, this is not feasible in practice, as DNS censorship is \cam{the most basic and prevalent form of censorship.} 
\cam{Censors often read plaintext DNS queries and interfere by blocking the query or injecting responses containing their own IP addresses~\cite{Hoang2021a, jin2021understanding}.}

There are \cam{several approaches for the client and proxy to select an optimized server IP address:}

\cam{(1) The client could perform an encrypted DNS request using methods like DNS over TLS (DoT) or DNS over HTTPS (DoH). The greatest challenge in using encrypted DNS is whether the client can access encrypted DNS resolvers, particularly since some countries are known to censor those resolvers~\cite{jin2021understanding}. There are several other considerations and limitations to using encrypted DNS for censorship circumvention. If the resolver is within the censor's regime, censors can manipulate the unencrypted traffic between the resolver and the nameserver~\cite{jin2021understanding}. Also, DoH downgrade to plaintext DNS is not uncommon in practice~\cite{lee2024measuring}.} 
(2) The proxy can leverage EDNS Client Subnet (ECS) to reveal the client's network prefix to the authoritative DNS server. 
\sjl{We note that anonymity is not a main concern for the client, as discussed in \secref{sec:threatmodel}.}
If the target domain’s authoritative DNS servers support ECS, the client will receive IP addresses that are closer in proximity to the client’s network specified in the DNS query. 
We note that ECS support is not ubiquitously available across recursive resolvers (e.g., Cloudflare~\cite{cfecs}) and authoritative servers. 
\textcolor{black}{The proxy can use a modified DNS client or run a local recursive resolver for DNS resolution, as standard DNS clients do not natively support ECS.} 
(3) The client may directly connect to the optimized server (or frontend) IP address obtained via an out-of-band channel (e.g., a system similar to Lox~\cite{lox} and rBridge~\cite{rbridge}). 

\begin{figure}[t]
    \centering
    \includegraphics[width=0.6\columnwidth]{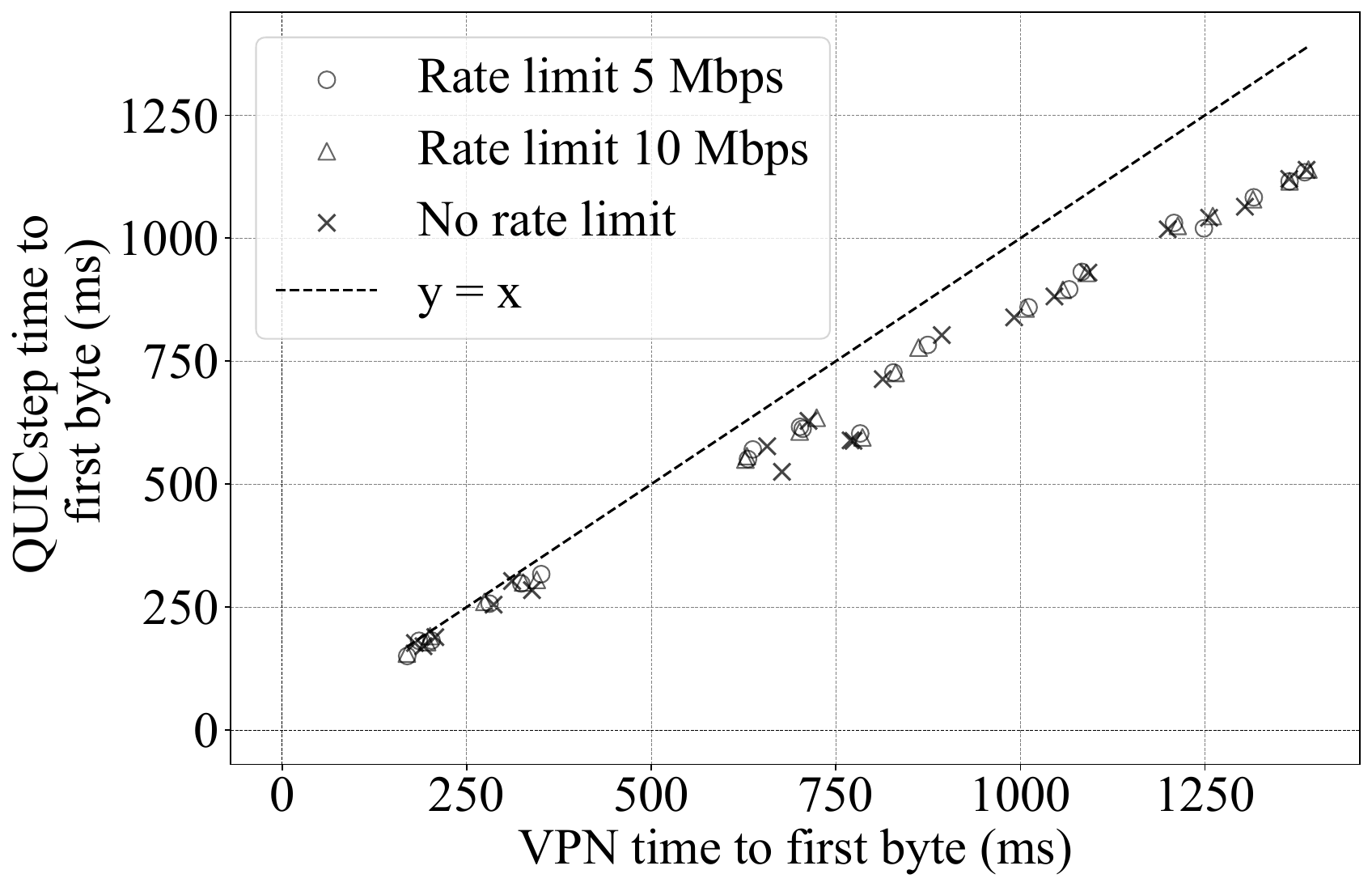}
    \caption{Time to first byte for VPN and \sysname connections when DNS is performed at the client. \sysname shows better time to first byte than the VPN connection unlike the default setting where the two values are comparable.}
    \label{fig:firstbyte_client}
\end{figure}

We evaluated \sysname's performance under the case where the client has some means of safely obtaining an optimized server IP address that is geographically close to itself.
All native, VPN, and \sysname connections used an IP address btained by the client, which eliminates the additional latency caused by the client connecting to a server far away from itself. 
Figure \ref{fig:firstbyte_client} shows that \sysname marginally reduces even the TTFB compared to VPN, unlike \figref{fig:firstbytes} where \sysname and VPN had comparable TTFB latencies. 
Table \ref{table:loadtime_dns_at_client} shows \sysname's page load time compared to VPN and native connections with different proxy locations. 
\sysname to VPN page load time ratio is significantly reduced for proxies far from the client. 
For example, with a proxy in Tokyo, using \sysname reduced latency by 32\% over VPN when the client in London accessed a server near Tokyo. But latency reduction was as great as 70\%, a gain 2.65 times more significant, when the client accessed a server near itself.

Again, it is important to note that while the DNS resolution process itself has manageable overhead, the primary factor is the server location determined by the DNS query location. 
By optimizing DNS resolution, \sysname performance can be boosted significantly.

\sjl{\paragraph{Summary of performance evaluations.} \sysname enables clients to make more efficient and effective use of secure \channelname{s} with limited performance by significantly reducing the amount of traffic that goes through the \channelname. \sysname's performance benefits are more pronounced when the \channelname incurs greater increase in page load time compared to the native channel: when the \channelname has low bandwidth and when the website is large. \sysname's performance is strongly correlated with the location of the proxy and the server. Using a proxy close to the client and connecting to a server close to the client by leveraging ECS or out-of-band channels can enhance \sysname performance.}

\begin{table}[t]
\centering
\begin{tabular}{|c|cc|cc|}
\hline
\multirow{2}{*}{Proxy} & \multicolumn{2}{c|}{\sysname/VPN} & \multicolumn{2}{c|}{\sysname/Native} \\ \cline{2-5} & \multicolumn{1}{l|}{Default} & \multicolumn{1}{l|}{\begin{tabular}[c]{@{}c@{}}DNS\\ optimized\end{tabular}} & \multicolumn{1}{l|}{Default} & \multicolumn{1}{l|}{\begin{tabular}[c]{@{}c@{}}DNS\\ optimized\end{tabular}} \\ \hline
\hline
Ireland & \multicolumn{1}{c|}{0.338} & 0.454 & \multicolumn{1}{c|}{1.066} & 1.000 \\
\multicolumn{1}{|l|}{Frankfurt} & \multicolumn{1}{c|}{0.340}   & 0.438 & \multicolumn{1}{c|}{1.328} & 0.997 \\
Montreal & \multicolumn{1}{c|}{0.463} & 0.423 & \multicolumn{1}{c|}{1.931} & 1.207 \\
Ohio & \multicolumn{1}{c|}{0.458} & 0.449 & \multicolumn{1}{c|}{1.847} & 1.175 \\
\textbf{Oregon} & \multicolumn{1}{c|}{\textbf{0.554}} & \textbf{0.238} & \multicolumn{1}{c|}{2.278} & 1.395 \\
\textbf{Seoul} & \multicolumn{1}{c|}{\textbf{0.586}} & \textbf{0.282} & \multicolumn{1}{c|}{4.920} & 2.949 \\
\textbf{Tokyo} & \multicolumn{1}{c|}{\textbf{0.682}} & \textbf{0.301} & \multicolumn{1}{c|}{3.842} & 2.444 \\ \hline
\end{tabular}
\caption{Ratio of QUICstep page load time to VPN and native connections when DNS resolution is optimized, compared to the default setup. Proxies are listed in order of distance from the client. The client is in London and the proxies' maximum throughput is 5\,Mbps. DNS optimization significantly enhances \sysname's performance gain over VPN when the proxy is far from the client (boldfaced).}
\label{table:loadtime_dns_at_client}
\end{table}

\section{Potential attacks against \sysname} \label{sec-analysis}

In this section, we explore potential attacks against \sysname and their feasibility, starting from simple protocol blocking and progressing to more sophisticated traffic analysis. 
We primarily focus on practical attacks that have been/could be employed by real-world censors, which are typically \emph{stateless} as studied in ~\cite{cat-and-mouse, Wu2023a, China-QUIC-Censorship}. 
We define a stateless attack as one that can be performed on a per-packet basis without requiring information from two or more packets. 
For example, QUIC SNI censorship can be stateless if the censor follows the reference implementation provided by Google~\cite{googlequicsni}.

\paragraph{DNS blocking and SNI censorship.} 
DNS blocking~\cite{Hoang2021a, Anonymous2020a} and SNI censorship are two major techniques censors employ to censor websites. 
\sysname evades DNS blocking and QUIC-SNI blocking as both DNS requests and QUIC \texttt{Handshake} packets are transmitted through a secure and encrypted tunnel that is not censored. 

\paragraph{IP address blocking.}
Some censors may adopt IP address blocking~\cite{Chai2019a}. As discussed in \secref{sec:threatmodel}, following prior work, \sysname is applicable to destinations whose IP addresses are not blocked. 
Censored domains can leverage non-blocked CDNs or cloud services to circumvent IP address blocking. 
Note that \sysname sometimes helps circumvent IP address-based blocking. 
In \sysname, DNS requests are performed through a proxy located outside of the censored regime and may resolve to a different IP address that may not be on the censor's IP address blocklist~\cite{Chai2019a}. 

\cam{The censor may also attempt to block the IP address of the \qsproxy. Handshake channel providers may remedy this by not making the IP addresses public like Tor bridges or frequently rotating IP addresses like Snowflake~\cite{snowflake}. The design of \sysname also helps reduce the risk of \qsproxy IP addresses being identified compared to relying on VPNs for all communication. One way to fingerprint VPN usage is identifying when the client communicates predominantly over a single IP address; this does not happen with \sysname as the client also communicates directly with the destination server.}


\paragraph{Blocking all QUIC traffic.}
One attack strategy to counter \sysname is to block all QUIC traffic. 
We argue that such an aggressive strategy would lead to high collateral damage, which could be undesirable for censors, given the rapid increase in QUIC/HTTP3 deployment. 
For example, major cloud providers in China (Alibaba, Tencent, Huawei, etc.) all provide QUIC-based services, and therefore blocking QUIC could negatively impact the revenue of these Chinese companies. 
We noticed that Russia was suspected of blocking QUIC traffic in 2022 during the early stage of QUIC deployment~\cite{xue2022tspu}. 
This is likely because DPIs could not decrypt SNIs in encrypted QUIC payloads at that time. 
However, as QUIC SNI detection techniques have matured and been integrated into commercial DPI systems (e.g., Cisco~\cite{ciscodpi}), it could incentivize nation-state censors to adopt QUIC SNI detection to minimize collateral damage. 
In fact, recent evidence indicates that Russia has transitioned to using QUIC SNI detection~\cite{russiasni}. 
As discussed, \sysname is an effective approach to bypass QUIC SNI censorship. 

\paragraph{Blocking all QUIC connection migrated traffic.} 
Theoretically, a censor could try to identify QUIC connection migration events and block all migrated QUIC connections. 
A potential indicator of a migrated connection is the absence of \texttt{Handshake} or \texttt{Initial} packets. 
However, such attacks face a significant challenge: \sysname migrations are indistinguishable from those triggered by typical client mobility events in terms of traffic characteristics. 
One can view \sysname as a system that uses artificial mobility events to trigger migration of native QUIC. 
There is no reliable way to differentiate between ``normal'' migration and \sysname migration on a per-connection basis. 
Connection migration is likely to become common in future mobile and vehicular networks, which is a key consideration that inspired the design of QUIC's connection migration feature. 
Therefore, simply blocking all migrated connections may cause substantial collateral damage. 

To evaluate the feasibility of censors dropping QUIC connections that do not contain \texttt{Handshake} or \texttt{Initial} packets, we analyzed QUIC traffic from a  campus network collected for 24 hours on November 6th, 2022.\footnote{The network traces are anonymized and we have obtained IRB approval from our institution. Refer to \secref{sec:ethics} for details.} 
\textbf{Out of 3,786,050 unique QUIC connections, 1,100,439 (29.1\%) did not contain a QUIC \texttt{Initial} nor a QUIC \texttt{Handshake} packet.} 
They are likely flows created from regular connection migration activity. 


\paragraph{Hypothetical case: Stateful traffic analysis.}
As a lightweight approach, \sysname does not consider stateful traffic analysis and makes no security claims against such attacks (e.g., detecting \sysname based on the abnormal frequency of connection migration events). 
Several deployed censorship circumvention tools also share a similar 
limitation, e.g., some Tor pluggable transports and VPNs can be fingerprinted via traffic analysis~\cite{wang2015seeing, xue2022openvpn, xue2024fingerprinting, wails2024precisely, almutairi2024fingerprinting}. 
We acknowledge that extensive research has been conducted on censorship circumvention techniques that are robust against stateful traffic analysis~\cite{brubaker2014cloudtransport, dyer2015marionette, barradas2017deltashaper, barradas2020poking, rosen2021balboa, figueira2022stegozoa, sun2023telepath, jia2023voiceover}. 
However, we note that such attacks are resource-intensive, as they require the censor to store significantly more state information. 
Practical real-time DPIs still favor lightweight detection mechanisms such as keyword matching (Section~\ref{sec:bk:censor}). 
It remains an open question whether advanced traffic analysis attacks against \sysname can achieve the efficiency required to meet real-time blocking goals of deployed censors.

\section{Discussion and conclusion} \label{sec:discussion}


In summary, \sysname presents a promising direction for censorship circumvention in a QUIC-first world.
\sysname successfully circumvents real-world censors and provides significant performance gain compared to relying on a secure but resource constrained channel for all communication.
\sysname is particularly useful in that it provides greater performance gain when the user fetches more data per-connection (e.g. when accessing large websites;~\secref{sec:eval:size}) and when the user has limited resources (e.g. when the \channelname has low bandwidth;~\secref{sec:eval:bandwidth}).

The application-agnostic nature of \sysname enables it to be integrated as part of existing censorship circumvention tools.
Integration of \sysname to existing proxy services can benefit providers as \sysname substantially reduces resource consumption of each client.
Our work also shows the current state of QUIC and connection migration support in the wild, and \sysname can also be used as a tool for evaluating connection migration support.

In this section, we discuss the path to large-scale \sysname deployment, for which the main bottleneck is the limited QUIC and connection migration support in the current Internet. 

\paragraph{Increasing QUIC support and adoption.} R{\"u}th et al. reported 1.2\% QUIC support among the Alexa top 1\,M list in October 2017~\cite{ruth2018first}; our results find over 22\% QUIC support among the Tranco top 1\,M list in November 2024.\footnote{The 2017 result did not exclude 404 error pages. Including 404 error pages, our numbers jump to around 30\%.} With the standardization of HTTP/3 and the ongoing increase in QUIC deployment we envision that QUIC will become the de facto norm for internet traffic~\cite{http3-cloudflare, http3-w3}.

\paragraph{Increasing connection migration support and adoption.}
Our results in \secref{sec:quicstat} demonstrate that connection migration support is increasing, but still does not extend to most QUIC-supporting websites. In our tests, we found \sysname is not only useful for censorship circumvention, but also for understanding whether and how websites support QUIC connection migration. In this work, we provide the most comprehensive measurement of QUIC connection migration deployment to-date, differentiating between different types of support. We hope our work will be a driver for increased connection migration support, and that service providers (e.g., Cloudflare) will recognize the power of connection migration.

\paragraph{Standardizing a mechanism for advertising connection migration support.} Due to the gap between QUIC and QUIC connection migration support as highlighted in our measurements, clients need to discover whether a host supports connection migration before they can leverage its performance benefits. Clients currently have no standard method for discovering connection migration support. As demonstrated by our own measurements in addition to prior work, it is not easy to reliably measure whether a website supports connection migration. In addition, many websites support only port or IP migration and not the other. The standardization of a method for discovering connection migration support, perhaps within an established QUIC channel, would allow all QUIC clients to benefit from the performance benefits of connection migration. At the moment, since it is nontrivial for clients to identify support for connection migration, clients need to maintain a list of endpoints that support connection migration.

\paragraph{Integrating \sysname into usable deployments.}
In this work, we demonstrate that implementing the core functionality \sysname is lightweight, and can be simplified to a number of packet-based routing rules. We suggest that with a mechanism for opportunistically discovering QUIC and connection migration support, \sysname can be deployed as a standalone tool (e.g. packaged as an Android VPN), or better yet, alongside existing censorship circumvention tooling. The performance benefits of \sysname also generalize to reducing bandwidth usage of a highly-secure \channelname, which could be leveraged by existing circumvention software to reduce load on volunteer (or otherwise limited) providers.

\section{Ethical considerations} \label{sec:ethics}

\paragraph{Censorship circumvention experiments.}
\textcolor{black}{Our censorship circumvention experiments in \secref{sec:real-world-censor} did not involve any human subjects. In our measurements with the GFW, we used a 
client machine (under our control) hosted at a large-scale commercial VPS provider with a dedicated IP address.
We accessed only a limited number of real-world domains to avoid the client machine itself being blocked. We ran a QUIC server under our control with different domain names for further testing. We note that this was run 
on localhost such that only our client machine (that knows the IP address of the QUIC server) could access it.}

\paragraph{Network trace analysis.}
To evaluate the collateral damage of blocking all QUIC connection migrated traffic in \secref{sec-analysis}, we analyzed anonymized traces gathered from our campus network.
\cam{Port 443 UDP traffic was captured by our campus network operator who managed the tap, sanitized the traces, anonymized IP addresses and source ports, and removed all protected payloads before releasing the traces to us.
The traces’ network storage was also managed by our campus network operator and protected with restricted access. These traces were used by other campus projects as well and we did not have control over the lifetime of the data.}
We obtained IRB approval from our institution to study this data and we did not perform any analysis beyond the aggregate statistic presented in the paper.

\begin{acks}
\cam{
We are grateful to anonymous reviewers for their helpful feedback.
This material is based on work supported by the Defense Advanced Research Project Agency (DARPA) under contract no. HR00112590081.
Any opinions, findings, conclusions, or recommendations expressed in this material are those of the authors, and do not necessarily reflect the views of the sponsors.}
\end{acks}

\bibliographystyle{plain}
\bibliography{references}

\begin{thebibliography}{10}

\bibitem{ooni}
{{OONI}: Open Observatory of Network Interference}.
\newblock In {\em 2nd USENIX Workshop on Free and Open Communications on the Internet (FOCI 12)}, Bellevue, WA, August 2012. USENIX Association.

\bibitem{almutairi2024fingerprinting}
Sultan Almutairi, Yogev Neumann, and Khaled Harfoush.
\newblock Fingerprinting vpns with custom router firmware: A new censorship threat model.
\newblock In {\em 2024 IEEE 21st Consumer Communications \& Networking Conference (CCNC)}, pages 976--981. IEEE, 2024.

\bibitem{cat-and-mouse}
Anonymous and Amonymous.
\newblock {Sharing a modified Shadowsocks as well as our thoughts on the cat-and-mouse game}, October 2022.

\bibitem{Anonymous2020a}
Anonymous, Arian~Akhavan Niaki, Nguyen~Phong Hoang, Phillipa Gill, and Amir Houmansadr.
\newblock Triplet censors: Demystifying {Great Firewall}'s {DNS} censorship behavior.
\newblock In {\em Free and Open Communications on the Internet}. USENIX, 2020.

\bibitem{barradas2017deltashaper}
Diogo Barradas, Nuno Santos, and Lu{\'\i}s Rodrigues.
\newblock {DeltaShaper: Enabling unobservable censorship-resistant TCP tunneling over videoconferencing streams}.
\newblock {\em Proceedings on Privacy Enhancing Technologies}, 2017(4):5--22, 2017.

\bibitem{barradas2020poking}
Diogo Barradas, Nuno Santos, Lu{\'\i}s Rodrigues, and V{\'\i}tor Nunes.
\newblock {Poking a hole in the wall: Efficient censorship-resistant Internet communications by parasitizing on WebRTC}.
\newblock In {\em Proceedings of the 2020 ACM SIGSAC Conference on Computer and Communications Security}, pages 35--48, 2020.

\bibitem{http3-cloudflare}
David Belson and Lucas Pardue.
\newblock Examining {HTTP/3} usage one year on, June 2023.

\bibitem{bishop2022rfc}
M~Bishop.
\newblock {RFC 9114: HTTP/3}, 2022.

\bibitem{bock2019geneva}
Kevin Bock, George Hughey, Xiao Qiang, and Dave Levin.
\newblock {Geneva: Evolving Censorship Evasion Strategies}.
\newblock In {\em Proceedings of the 2019 ACM SIGSAC Conference on Computer and Communications Security}, CCS '19, page 2199–2214, New York, NY, USA, 2019. Association for Computing Machinery.

\bibitem{2020Bock}
Kevin Bock, iyouport, Anonymous, Louis-Henri Merino, David Fifield, Amir Houmansadr, and Dave Levin.
\newblock Exposing and circumventing china's censorship of esni.
\newblock Technical report, GFW Report, August 2020.

\bibitem{snowflake}
Cecylia Bocovich, Arlo Breault, David Fifield, Serene, and Xiaokang Wang.
\newblock Snowflake, a censorship circumvention system using temporary {WebRTC} proxies.
\newblock In {\em 33rd USENIX Security Symposium (USENIX Security 24)}, pages 2635--2652, Philadelphia, PA, August 2024. USENIX Association.

\bibitem{brubaker2014cloudtransport}
Chad Brubaker, Amir Houmansadr, and Vitaly Shmatikov.
\newblock Cloudtransport: Using cloud storage for censorship-resistant networking.
\newblock In {\em International Symposium on Privacy Enhancing Technologies Symposium}, pages 1--20. Springer, 2014.

\bibitem{buchet2024analysisquicconnectionmigration}
Aurélien Buchet and Cristel Pelsser.
\newblock An analysis of {QUIC} connection migration in the wild, 2024.

\bibitem{callanan2011leaping}
Cormac Callanan, Hein Dries-Ziekenheiner, Alberto Escudero-Pascual, and Robert Guerra.
\newblock Leaping over the firewall: A review of censorship circumvention tools.
\newblock {\em Report by Freedom House}, 2011.

\bibitem{Chai2019a}
Zimo Chai, Amirhossein Ghafari, and Amir Houmansadr.
\newblock On the importance of encrypted-{SNI} ({ESNI}) to censorship circumvention.
\newblock In {\em Free and Open Communications on the Internet}. USENIX, 2019.

\bibitem{ciscodpi}
Cisco.
\newblock {Support for SNI Detection}, unknown.

\bibitem{cfecs}
Cloudflare.
\newblock {1.1.1.1 (DNS Resolver) FAQ}, 2024.

\bibitem{devraj2021redact}
Arjun Devraj, Liang Wang, and Jennifer Rexford.
\newblock Redact: refraction networking from the data center.
\newblock {\em ACM SIGCOMM Computer Communication Review}, 51(4):15--22, 2021.

\bibitem{dyer2015marionette}
Kevin~P Dyer, Scott~E Coull, and Thomas Shrimpton.
\newblock Marionette: A programmable network traffic obfuscation system.
\newblock In {\em 24th $\{$USENIX$\}$ Security Symposium ($\{$USENIX$\}$ Security 15)}, pages 367--382, 2015.

\bibitem{ebpf}
The eBPF Foundation.
\newblock {eBPF Documentation}, 2024.

\bibitem{elmenhorst2022blog}
Kathrin Elmenhorst.
\newblock {A Quick Look at QUIC Censorship}, Apr 2022.

\bibitem{elmenhorst2021web}
Kathrin Elmenhorst, Bertram Sch{\"u}tz, Nils Aschenbruck, and Simone Basso.
\newblock {Web censorship measurements of HTTP/3 over QUIC}.
\newblock In {\em Proceedings of the 21st ACM Internet Measurement Conference}, pages 276--282, 2021.

\bibitem{fifield2015blocking}
David Fifield, Chang Lan, Rod Hynes, Percy Wegmann, and Vern Paxson.
\newblock Blocking-resistant communication through domain fronting.
\newblock {\em Proceedings on Privacy Enhancing Technologies}, 2015.

\bibitem{figueira2022stegozoa}
Gabriel Figueira, Diogo Barradas, and Nuno Santos.
\newblock {Stegozoa: Enhancing WebRTC Covert Channels with Video Steganography for Internet Censorship Circumvention}.
\newblock In {\em Proceedings of the 2022 ACM on Asia Conference on Computer and Communications Security}, pages 1154--1167, 2022.

\bibitem{googlequicsni}
Google.
\newblock {Parsing QUIC Client Hellos}, 2021.

\bibitem{quiche}
Google.
\newblock {QUICHE}.
\newblock \url{https://quiche.googlesource.com/quiche/}, 2022.

\bibitem{govil2020mimiq}
Yashodhar Govil, Liang Wang, and Jennifer Rexford.
\newblock {MIMIQ: Masking IPs with migration in QUIC}.
\newblock In {\em 10th USENIX Workshop on Free and Open Communications on the Internet (FOCI)}, 2020.

\bibitem{Hoang2024a}
Nguyen~Phong Hoang, Jakub Dalek, Masashi Crete-Nishihata, Nicolas Christin, Vinod Yegneswaran, Michalis Polychronakis, and Nick Feamster.
\newblock {GFWeb}: Measuring the {Great Firewall}'s {Web} censorship at scale.
\newblock In {\em USENIX Security Symposium}. USENIX, 2024.

\bibitem{Hoang2021a}
Nguyen~Phong Hoang, Arian~Akhavan Niaki, Jakub Dalek, Jeffrey Knockel, Pellaeon Lin, Bill Marczak, Masashi Crete-Nishihata, Phillipa Gill, and Michalis Polychronakis.
\newblock How great is the {Great Firewall}? {Measuring} {China}'s {DNS} censorship.
\newblock In {\em USENIX Security Symposium}. USENIX, 2021.

\bibitem{houmansadr2013freewave}
Amir Houmansadr, Thomas~J Riedl, Nikita Borisov, and Andrew~C Singer.
\newblock I want my voice to be heard: {IP} over {Voice-over-IP} for unobservable censorship circumvention.
\newblock In {\em NDSS}, 2013.

\bibitem{hysteria}
{Hysteria developers}.
\newblock {Hysteria}.

\bibitem{iyengar2021rfc}
Jana Iyengar and Martin Thomson.
\newblock {RFC 9000: QUIC: A UDP-Based Multiplexed and Secure Transport}.
\newblock {\em Omtermet Emgomeeromg Task Force}, 2021.

\bibitem{jia2023voiceover}
Watson Jia, Joseph Eichenhofer, Liang Wang, and Prateek Mittal.
\newblock Voiceover: Censorship-circumventing protocol tunnels with generative modeling.
\newblock {\em Free and Open Communications on the Internet}, 2023.

\bibitem{jin2021understanding}
Lin Jin, Shuai Hao, Haining Wang, and Chase Cotton.
\newblock Understanding the impact of encrypted {DNS} on internet censorship.
\newblock In {\em Proceedings of the Web Conference 2021}, pages 484--495, 2021.

\bibitem{eksisozluk}
Sedat Kappanoğlu.
\newblock turkish {ISPs} use two methods for blocking access...
\newblock \url{https://twitter.com/esesci/status/1630024112071491586}.

\bibitem{lanternlimit}
Lantern.
\newblock {Lantern - Frequently Asked Questions}.

\bibitem{LePochat2019}
Victor {Le Pochat}, Tom {Van Goethem}, Samaneh Tajalizadehkhoob, Maciej Korczy\'{n}ski, and Wouter Joosen.
\newblock Tranco: A research-oriented top sites ranking hardened against manipulation.
\newblock In {\em Proceedings of the 26th Annual Network and Distributed System Security Symposium}, NDSS 2019, February 2019.

\bibitem{lee2024measuring}
Jinseo Lee, David Mohaisen, and Min~Suk Kang.
\newblock Measuring dns-over-https downgrades: Prevalence, techniques, and bypass strategies.
\newblock {\em Proceedings of the ACM on Networking}, 2(CoNEXT4):1--22, 2024.

\bibitem{manfredi2018multiflow}
Victoria Manfredi and Pi~Songkuntham.
\newblock {MultiFlow}: {Cross-Connection} decoy routing using {TLS} 1.3 session resumption.
\newblock In {\em 8th USENIX Workshop on Free and Open Communications on the Internet (FOCI 18)}, Baltimore, MD, August 2018. USENIX Association.

\bibitem{mohajeri2012skypemorph}
Hooman Mohajeri~Moghaddam, Baiyu Li, Mohammad Derakhshani, and Ian Goldberg.
\newblock {Skypemorph: Protocol obfuscation for tor bridges}.
\newblock In {\em Proceedings of the 2012 ACM conference on Computer and communications security}, pages 97--108, 2012.

\bibitem{nasr2020massbrowser}
Milad Nasr, Hadi Zolfaghari, Amir Houmansadr, and Amirhossein Ghafari.
\newblock Massbrowser: Unblocking the censored web for the masses, by the masses.
\newblock In {\em NDSS}, 2020.

\bibitem{niere2025encrypted}
Niklas Niere, Felix Lange, Nico Heitmann, and Juraj Somorovsky.
\newblock Encrypted client hello (ech) in censorship circumvention.
\newblock {\em Free and Open Communications on the Internet}, 2025.

\bibitem{meek}
Tor Project.
\newblock meek, 2020.

\bibitem{rfc8446}
Eric Rescorla.
\newblock {The Transport Layer Security (TLS) Protocol Version 1.3}.
\newblock RFC 8446, August 2018.

\bibitem{ech-25}
Eric Rescorla, Kazuho Oku, Nick Sullivan, and Christopher~A. Wood.
\newblock {TLS Encrypted Client Hello}.
\newblock Internet-Draft draft-ietf-tls-esni-25, Internet Engineering Task Force, June 2025.
\newblock Work in Progress.

\bibitem{rosen2021balboa}
Marc~B Rosen, James Parker, and Alex~J Malozemoff.
\newblock Balboa: Bobbing and weaving around network censorship.
\newblock In {\em 30th USENIX Security Symposium (USENIX Security 21)}, pages 3399--3413, 2021.

\bibitem{accessnow}
Zach Rosson, Felicia, Carolyn Tackett, and Meabh Maguire.
\newblock Lives on hold: internet shutdowns in 2024, February 2025.

\bibitem{shadowsocks-rust}
Shadowsocks rust developers.
\newblock Shadowsocks-rust.

\bibitem{ruth2018first}
Jan R{\"u}th, Ingmar Poese, Christoph Dietzel, and Oliver Hohlfeld.
\newblock A first look at quic in the wild.
\newblock In {\em Passive and Active Measurement: 19th International Conference, PAM 2018, Berlin, Germany, March 26--27, 2018, Proceedings 19}, pages 255--268. Springer, 2018.

\bibitem{Satija2021a}
Sambhav Satija and Rahul Chatterjee.
\newblock {BlindTLS}: Circumventing {TLS}-based {HTTPS} censorship.
\newblock In {\em Free and Open Communications on the Internet}. ACM, 2021.

\bibitem{shadowsocks-python}
{Shadowsocks developers}.
\newblock {Shadowsocks}.

\bibitem{sing-box}
{Sing-box developers}.
\newblock {Sing-box}.

\bibitem{sun2023telepath}
Zhen Sun and Vitaly Shmatikov.
\newblock Telepath: A minecraft-based covert communication system.
\newblock In {\em 2023 IEEE Symposium on Security and Privacy (SP)}, pages 2223--2237. IEEE, 2023.

\bibitem{censored_planet}
Ram Sundara~Raman, Prerana Shenoy, Katharina Kohls, and Roya Ensafi.
\newblock {Censored Planet: An Internet-Wide, Longitudinal Censorship Observatory}.
\newblock In {\em Proceedings of the 2020 ACM SIGSAC Conference on Computer and Communications Security}, CCS '20, page 49–66, New York, NY, USA, 2020. Association for Computing Machinery.

\bibitem{sundararaman2020measuring}
Ram Sundara~Raman, Adrian Stoll, Jakub Dalek, Reethika Ramesh, Will Scott, and Roya Ensafi.
\newblock {Measuring the Deployment of Network Censorship Filters at Global Scale.}
\newblock In {\em NDSS}, 2020.

\bibitem{torthroughput}
{The Tor Project}.
\newblock Tor metrics - performance, 2024.

\bibitem{rfc9001}
Martin Thomson and Sean Turner.
\newblock {Using TLS to Secure QUIC}.
\newblock RFC 9001, May 2021.

\bibitem{lox}
Lindsey Tulloch.
\newblock Lox: Protecting the social graph in bridge distribution.
\newblock Master's thesis, University of Waterloo, 2022.

\bibitem{v2ray}
{V2Ray developers}.
\newblock {V2Ray}.

\bibitem{Vines2024c}
Paul Vines, Samuel McKay, Jesse Jenter, and Suresh Krishnaswamy.
\newblock Communication breakdown: Modularizing application tunneling for signaling around censorship.
\newblock {\em Privacy Enhancing Technologies}, 2024(1), 2024.

\bibitem{http3-w3}
W3Techs.
\newblock Usage statistics of http/3 for websites, May 2025.

\bibitem{wails2024precisely}
Ryan Wails, George~Arnold Sullivan, Micah Sherr, and Rob Jansen.
\newblock On precisely detecting censorship circumvention in real-world networks.
\newblock In {\em Network and Distributed System Security}, 2024.

\bibitem{wang2015seeing}
Liang Wang, Kevin~P Dyer, Aditya Akella, Thomas Ristenpart, and Thomas Shrimpton.
\newblock Seeing through network-protocol obfuscation.
\newblock In {\em Proceedings of the 22nd ACM SIGSAC Conference on Computer and Communications Security}, pages 57--69, 2015.

\bibitem{wang2022leveraging}
Mona Wang, Anunay Kulshrestha, Liang Wang, and Prateek Mittal.
\newblock Leveraging strategic connection migration-powered traffic splitting for privacy.
\newblock In {\em Proceedings on Privacy Enhancing Technologies}, page 498–515, 2022.

\bibitem{rbridge}
Qiyan Wang, Zi~Lin, Nikita Borisov, and Nicholas Hopper.
\newblock rbridge: User reputation based tor bridge distribution with privacy preservation.
\newblock In {\em NDSS}, 2013.

\bibitem{psiphonthroughput2}
Jon Watson.
\newblock {How to Use Psiphon The Censorship-Circumvention Tool}, 2021.

\bibitem{domain-shadowing}
Mingkui Wei.
\newblock Domain shadowing: Leveraging content delivery networks for robust {Blocking-Resistant} communications.
\newblock In {\em 30th USENIX Security Symposium (USENIX Security 21)}, pages 3327--3343. USENIX Association, August 2021.

\bibitem{psiphonthroughput1}
Mike Williams.
\newblock Pshiphon review, 2020.

\bibitem{russiasni}
wkrp.
\newblock {Throttling→blocking of YouTube in Russia, 2024-07-12}, 2024.

\bibitem{Wu2023a}
Mingshi Wu, Jackson Sippe, Danesh Sivakumar, Jack Burg, Peter Anderson, Xiaokang Wang, Kevin Bock, Amir Houmansadr, Dave Levin, and Eric Wustrow.
\newblock How the {Great Firewall} of {China} detects and blocks fully encrypted traffic.
\newblock In {\em USENIX Security Symposium}. USENIX, 2023.

\bibitem{xray}
{XRay developers}.
\newblock {XRay}.

\bibitem{xue2024bridging}
Diwen Xue, Anna Ablove, Reethika Ramesh, Grace~Kwak Danciu, and Roya Ensafi.
\newblock Bridging barriers: A survey of challenges and priorities in the censorship circumvention landscape.
\newblock In {\em 33rd USENIX Security Symposium (USENIX Security 24)}, pages 2671--2688, Philadelphia, PA, August 2024. USENIX Association.

\bibitem{xue2024fingerprinting}
Diwen Xue, Michalis Kallitsis, Amir Houmansadr, and Roya Ensafi.
\newblock Fingerprinting obfuscated proxy traffic with encapsulated $\{$TLS$\}$ handshakes.
\newblock In {\em 33rd USENIX Security Symposium (USENIX Security 24)}, pages 2689--2706, 2024.

\bibitem{xue2022tspu}
Diwen Xue, Benjamin Mixon-Baca, ValdikSS, Anna Ablove, Beau Kujath, Jedidiah~R Crandall, and Roya Ensafi.
\newblock Tspu: Russia's decentralized censorship system.
\newblock In {\em Proceedings of the 22nd ACM Internet Measurement Conference}, pages 179--194, 2022.

\bibitem{xue2022openvpn}
Diwen Xue, Reethika Ramesh, Arham Jain, Michaelis Kallitsis, J~Alex Halderman, Jedidiah~R Crandall, and Roya Ensafi.
\newblock {OpenVPN is open to VPN fingerprinting}.
\newblock {\em Communications of the ACM}, 2022.

\bibitem{China-QUIC-Censorship}
Ali Zohaib, Qiang Zao, Jackson Sippe, Abdulrahman Alaraj, Amir Houmansadr, Zakir Durumeric, and Eric Wustrow.
\newblock Exposing and circumventing {SNI}-based {QUIC} censorship of the {Great Firewall of China}.
\newblock In {\em USENIX Security Symposium}. USENIX, 2025.

\end{thebibliography}

\appendix

\section{Additional performance evaluations}

\begin{table*}[t]
\centering
\begin{tabular}{|c|c|c|c|c|c|}
\hline
 \textbf{Client} & \textbf{Proxy} & \multicolumn{1}{c|}{\textbf{\begin{tabular}[c]{@{}c@{}}Max throughput\\ 1 Mbps\end{tabular}}} & \multicolumn{1}{c|}{\textbf{\begin{tabular}[c]{@{}c@{}}Max throughput\\ 5 Mbps\end{tabular}}} & \multicolumn{1}{c|}{\textbf{\begin{tabular}[c]{@{}c@{}}Max throughput\\ 10 Mbps\end{tabular}}} & \textbf{No rate limit} \\ 
 \hline
 \hline
  & Ireland & 0.069 & 0.338 & 0.510 & 0.996 \\
  & Frankfurt & 0.070 & 0.340 & 0.616 & 0.898 \\
  & Montreal & 0.072 & 0.463 & 0.738 & 1.052 \\
 \textbf{London} & Ohio & 0.088 & 0.458 & 0.588 & 0.996 \\
  & Oregon & 0.088 & 0.554 & 0.860 & 1.071 \\
  & Seoul & 0.128 & 0.586 & 0.633 & 0.765 \\
  & Tokyo & 0.124 & 0.682 & 0.976 & 1.553 \\
 \hline
  & Tokyo & 0.079 & 0.269 & 0.449 & 0.971 \\
  & Seoul & 0.079 & 0.319 & 0.511 & 0.868 \\
  & Oregon & 0.095 & 0.476 & 0.740 & 1.027 \\
 \textbf{Osaka} & Frankfurt & 0.255 & 0.719 & 1.015 & 1.273 \\
  & Ireland & 0.223 & 0.811 & 1.293 & 1.183 \\
  & Montreal & 0.136 & 0.560 & 0.752 & 1.002 \\
  & Ohio & 0.134 & 0.519 & 0.751 & 0.965 \\
 \hline
  & Montreal & 0.140 & 0.876 & 0.956 & 0.963  \\ 
  & Ohio &0.147 & 0.882 & 0.977 & 0.973 \\
  & Oregon & 0.156 & 0.895 & 0.932 & 1.010 \\
 \textbf{New Jersey} & Ireland & 0.210 & 0.921 & 1.194 & 1.065 \\
  & Frankfurt & 0.213 & 0.952 & 1.216 & 1.174 \\
  & Tokyo & 0.236 & 0.951 & 1.032 & 1.117 \\
  & Seoul & 0.265 & 0.750 & 0.834 & 0.776 \\
 \hline
\end{tabular}
\caption{Ratio of \sysname page load time to VPN page load time with different client locations, proxy locations, and proxy rate limits. For each client location the proxies are listed in order of geographical distance from the client.}
\label{table:loadtime_throughput_full}
\end{table*}

\begin{table*}[]
\centering
\begin{tabular}{|c|c|c|c|c|c|c|c|}
\hline
 \multicolumn{1}{|c|}{\textbf{\begin{tabular}[c]{@{}c@{}}Rate \\ limit\end{tabular}}} & \multicolumn{1}{c|}{\textbf{\begin{tabular}[c]{@{}c@{}}Website \\ size\end{tabular}}} & \textbf{Native (ms)} & \textbf{VPN (ms)} & \textbf{\sysname (ms)} & \textbf{\sysname\,/\,VPN} & \textbf{\sysname\,/\,Native} & \begin{tabular}[c]{@{}c@{}}\textbf{\sysname\,-}\\ \textbf{Native (ms)}\end{tabular} \\ 
 \hline
 \hline
 {} & 10\,KB & 98.97 & 926.26 & 790.41 & 0.853 & 7.986 & 691.44 \\
 5\,Mbps & 100\,KB & 145.11 & 1411.53 & 837.77 & 0.594 & 5.773 & 692.66 \\
 {} & 1\,MB & 277.4 & 3128.94 & 971.22 & 0.310 & 3.501 & 693.82 \\
 {} & 10\,MB & 1272.28 & 18894.73 & 1962.12 & 0.104 & 1.542 & 689.84 \\
 \hline
 {} & 10\,KB & 97.22 & 928.26 & 791.27 & 0.852 & 8.139 & 694.05 \\
 10\,Mbps & 100\,KB & 138.41 & 1409.3 & 830.78 & 0.589 & 6.002 & 692.37 \\
 {} & 1\,MB & 262.29 & 2559.01 & 956.57 & 0.374 & 3.647 & 694.28 \\
 {} & 10\,MB & 1040.3 & 10431.35 & 1737.8 & 0.167 & 1.670 & 697.50 \\
 \hline
 {} & 10\,KB & 104.99 & 924.11 & 790.55 & 0.855 & 7.530 & 685.56 \\
 None & 100\,KB & 157.57 & 1404.99 & 844.25 & 0.601 & 5.358 & 686.68 \\
 {} & 1\,MB & 351.62 & 2413.92 & 1038.42 & 0.430 & 2.953 & 686.80 \\
 {} & 10\,MB & 999.588 & 3875.32 & 1674.96 & 0.432 & 1.676 & 675.37 \\
 \hline
\end{tabular}
\caption{Latency (ms) with different rate limits and website sizes.}
\label{table:size}
\end{table*}

\end{document}